\documentclass[10pt]{iopart}

\usepackage{iopams}
\expandafter\let\csname equation*\endcsname\relax

\expandafter\let\csname endequation*\endcsname\relax

\usepackage{amsmath}

\usepackage{stmaryrd} 

\usepackage{MnSymbol} 
\usepackage{csquotes}
\usepackage{hyperref} 
\hypersetup{%
    colorlinks=true,
    linkcolor=blue,
    urlcolor=magenta
}
\usepackage[english]{babel}
\usepackage{amssymb}
\usepackage{bm}
\usepackage{longtable}
\usepackage{booktabs}
\usepackage{graphicx}
\usepackage{xcolor} 

\usepackage{doi}

\definecolor{light-gray}{gray}{0.95}

\newcommand{\eps}{\varepsilon}
\renewcommand{\d}{\mathrm{d}}
\renewcommand{\vec}[1]{{\bm{#1}}}

\newcommand{\vol}{\text{vol}}
\newcommand{\dA}{\,\mathrm{dA}}
\newcommand{\dV}{\mathrm{dV}\,}

\newcommand{\ExB}{$\bm{E}\times\bm{B}\ $}

\newcommand{\RA}[1]{\left \langle #1 \right \rangle}
\newcommand{\FA}[1]{\left\llbracket #1 \right\rrbracket}
\newcommand{\FMA}[1]{\left\llbracket #1 \right\rrbracket_M}
\newcommand{\RF}[1]{\widetilde{#1}}
\newcommand{\FF}[1]{\widehat{#1}}

\newcommand{\sumsp}{\sum_{\mathrm{s}} }
\newcommand{\Pem}{ P_{\mathrm{em}} }
\newcommand{\vPem}{ \vec{P}_{\mathrm{em}} }
\newcommand{\Mem}{ {M^{\mathrm{em}}_\perp} }
\newcommand{\vMem}{\vec{ M}^{\mathrm{em}}_\perp }

\newcommand{\vPgy}{ \vec{P}_{\mathrm{gy}} }

\newcommand{\vMgy}{\vec{ M}^{\mathrm{gy}} }

\newcommand{\form}{{\mathcal I }}
\newcommand{\urho}{{\vec L}}

\newcommand{\bhat}{\bm{\hat{b}}}
\newcommand{\bperp}{\widetilde{ \vec b_\perp}}
\newcommand{\rhohat}{\bm{\hat\rho}}
\newcommand{\ehat}{\bm{\hat{e}}}
\newcommand{\ephi}{\bm{ e_\varphi}}
\newcommand{\etheta}{\bm{ e_\vartheta}}
\newcommand{\eeta}{\bm{ e_\eta}}

\newcommand{\np}{\vec\nabla_{\perp}}
\newcommand{\nc}{\vec\nabla\cdot}
\newcommand{\cn}{\cdot\vec\nabla}
\newcommand{\vn}{\vec{\nabla}}
\newcommand{\npar}{\nabla_\parallel}

\newcommand{\T}{\mathrm{T}}
\newcommand{\tv}{ {\; v} }

\newcommand{\highlight}[1]{#1}

\newcommand{\LWL}{long-wavelength limit}
\newcommand{\DK}{drift }

\newcommand{\TotalStress}[2]{\ensuremath{\mathcal{T}_{#1}^{#2}}} 
\newcommand{\Tstress}[2]{\ensuremath{\mathcal{F}_{#1}^{#2}}} 
\newcommand{\FExBstress}[2]{\ensuremath{\mathcal{F}_{E,#1}^{#2}}} 
\newcommand{\Fdiastress}[2]{\ensuremath{\mathcal{F}_{D,#1}^{#2}}} 

\newcommand{\Xstress}[2]{\ensuremath{\mathcal{F}_{F,#1}^{#2}}} 
\newcommand{\Hstress}[2]{\ensuremath{\mathcal{F}_{F, #1}^{#2}}} 
\newcommand{\Kstress}[2]{\ensuremath{\mathcal{K}_{#1}^{#2}}} 
\newcommand{\Mstress}[2]{\ensuremath{\mathcal{M}_{#1}^{#2}}} 
\newcommand{\MBstress}[2]{\ensuremath{\mathcal{M}_{B,#1}^{#2}}} 
\newcommand{\Wstress}[2]{\ensuremath{\mathcal{M}_{M,#1}^{#2}}} 
\newcommand{\Rstress}[2]{\ensuremath{\mathcal{R}_{E,#1}^{#2}}} 
\newcommand{\Dstress}[2]{\ensuremath{\mathcal{R}_{D,#1}^{#2}}} 

\begin{document}

\title{Angular momentum and rotational energy of mean flows in toroidal magnetic fields}
\author{M. Wiesenberger$^1$ and M. Held$^2$}
\address{$^1$ Department of Physics, Technical University of Denmark, DK-2800 Kgs. Lyngby, Denmark}
\address{$^2$ Department of Space, Earth and Environment, Chalmers University of Technology, SE-412 96 Gothenburg, Sweden}
\ead{mattwi@fysik.dtu.dk}
\begin{abstract}
We derive the balance equation for the Favre averaged angular momentum in toroidal not necessarily axisymmetric magnetic field equilibria. 
We find that the components of angular momentum are given by the covariant poloidal and toroidal components of \ExB and parallel flow velocities
and we separately identify all relevant stress tensors, torques and source terms for each of these components.
Our results feature the Favre stress generalisations of previously found Reynolds stresses like the diamagnetic or parallel \ExB stress, as well as the density gradient drive term. Further, we identify the magnetic shear as a source of poloidal \ExB angular momentum and discuss the mirror and the Lorentz force. Here, we find that the geodesic transfer term, the Stringer-Winsor spin-up term and the ion-orbit loss term are all part of the Lorentz force and are in fact one and the same term. 

Discussing the relation to angular velocity we build the inertia tensor with the help of the first fundamental form of a flux-surface. In turn, the inertia tensor is used to construct a flux-surface averaged rotational energy for \ExB surface flows of the plasma. The evolution of this rotational energy features a correction of previous results due to the inertia tensor. In particular, this correction suggests that density sources on the high-field side contribute much more to zonal flow energy generation than on the low field side.

Our derivation is based on a full-F, electromagnetic, gyro-kinetic model in a \LWL. The results can be applied to gyro-kinetic as well as gyro-fluid theories and can also be compared to drift-kinetic and drift-fluid models. Simplified cases for the magnetic field geometry including the axisymmetric purely toroidal and purely poloidal magnetic fields are discussed, as are the angular momentum balance of the electromagnetic fields, the ion-orbit loss mechanism and the parallel acceleration.
\end{abstract}


\submitto{Nuclear Fusion}
\ioptwocol

\section{Introduction}
The double periodicity of a toroidal magnetic field configuration can be associated with two rotational degrees of freedom: toroidal and poloidal rotation. In a toroidally confined plasma both toroidal and poloidal rotation are observed and subject to intensive research.

Studies of toroidal rotation favour the toroidally symmetric tokamak case, where the symmetry leads to the exact conservation of the collective\footnote{after species and particle summation - individual particles exchange momentum through fluctuating electromagnetic fields} canonical angular momentum~\cite{Scott2010,Brizard2011,Stoltzfus2012}.
Of particular interest is the so-called intrinsic rotation, which refers
to the ability of the plasma to spontaneously rotate without application of
an external torque like neutral beam injection~\cite{Rice2007,Diamond2013,Stoltzfus2019}.
This is an important topic because toroidal rotation stabilizes the plasma against instabilities like the resistive wall mode. 

The ideal toroidal symmetry of a tokamak is broken in stellarators and in reality also in tokamaks due to magnetic ripple effects from external field coils spacing.
In fact, stellarator physics is different from tokamaks in some important aspects~\cite{Helander2012}. Neoclassical transport levels are much higher in a stellarator than in a tokamak even though stellarator optimization aims at reducing these levels down or below turbulent transport levels. More importantly however, the exact invariance of toroidal angular momentum is lost in a stellarator due to the lack of axial symmetry\footnote{Axisymmery, axial symmetry and toroidal symmetry are used interchangeably throughout this manuscript.}. It is argued that in this case it is impossible for the plasma to rotate as fast as in (quasi-)axisymmetric devices~\cite{Helander2008, Sugama2011} since the radial electric field is restricted by the ambipolarity condition but that zonal flows may still develop.

Poloidal angular momentum, just as toroidal angular momentum, has two components in a general magnetic field, one stemming from the parallel velocity projected to the poloidal direction $u_\parallel \bhat \cdot \etheta$, the other from the drifts
perpendicular to the magnetic field $\vec u_\perp \cdot \etheta$ (toroidal momentum analogously with $\ephi$). Here, $u_\parallel \equiv \vec u\cdot\bhat$ is the parallel flow velocity, $\vec u_\perp\equiv \bhat\times(\vec u\times \bhat)$ is the perpendicular flow velocity, $\bhat$ is the magnetic unit vector and $\etheta$ and $\ephi$ are the covariant poloidal and toroidal base vectors.
In reverse this means that both parallel velocity as well as the perpendicular drifts contribute to both toroidal as well as poloidal rotation. 
This is simply the geometrical observation that parallel and perpendicular directions versus poloidal and toroidal directions are different basis vectors for a flux-surface.
This being said, the poloidal component of \ExB velocity $u_{E,\vartheta}\equiv \vec u_E\cdot\etheta$ gains significant interest
 because of its role in
the formation of a transport barrier during the L-H transition~\cite{Diamond2005,Fujisawa2009,Gurcan2015}.
The high confinement mode is accompanied by a narrow potential well just inside the separatrix of a diverted magnetic field geometry. The associated radial electric field drives a strongly sheared and flux-aligned \ExB mean flow, which suppresses turbulence and thus reduces the radial flow of particles and heat out of the confined region. This
\ExB shear flow is believed to emerge out of turbulent fluctuations via the Reynolds stress, yet other mechanisms like the ion-orbit loss mechanism~\cite{Connor2000,Chang2002,Ku2018} or the Favre stress and background density gradient drive~\cite{Held2018} are currently under discussion as well.
Recent results suggest that the latter significantly alter the generation mechanism of \ExB zonal flows for high density fluctuation amplitudes and steep density gradients~\cite{Held2018,Held2019}.

It is instructive to introduce rotation also from a purely mechanical perspective. Consider a particle of mass $m$ confined to a toroidal surface. 
Its Lagrangian reads $L_p = m(R^2 \dot\varphi^2 + a^2\dot \vartheta^2)/2$ with the geometrical toroidal angle $\varphi$ and poloidal angle $\vartheta$. In
an ideal torus the distance from the major axis $R(\vartheta) = R_0 + a\cos \vartheta$, with $R_0$ the major radius, is
independent of the geometric toroidal angle $\varphi$. The distance from the minor axis $a=a_0$ remains the minor radius $a_0$. 
The Euler-Lagrange equations directly yield the conservation of toroidal angular momentum
$\dot L_\varphi = 0$ with $L_\varphi = m R^2 \dot \varphi$. This is a consequence of 
the independence of $R$ and $a$ of the toroidal angle $\varphi$. We then have $\dot \varphi = L_\varphi/m (R_0 + a\cos\vartheta)^2$. Notice that the angular frequency $\dot\varphi$ is higher on the torus inside $\vartheta=\pi$ than on the outside $\vartheta = 0$, which we intuitively expect. 
In contrast, the equation for the poloidal angle is given by 
the nonlinear differential equation $\ddot\vartheta = - L_\varphi^2\sin\vartheta/ m^2 a^4( R_0/a + \cos\vartheta)^3 $.
We observe that the poloidal angular momentum is not a conserved quantity for $L_\varphi \neq 0$. Furthermore,  on a generally shaped toroidal flux-surface like that of a stellarator $R$ as well as $a$ depend on both $\varphi$ and $\theta$. There, neither toroidal nor poloidal angular momenta are conserved and $\vartheta$ and $\varphi$ obey a coupled set of nonlinear differential equations.

In this contribution we calculate the toroidal and poloidal angular momentum balance separately for both the \ExB and the parallel velocity part.
Previous work is mostly restricted to toroidal symmetry~\cite{Scott2010,Brizard2011,Stoltzfus2012}, simplified magnetic field
geometry~\cite{Scott2003,Naulin2005,Madsen2017,Held2018} or delta-f modelling~\cite{Scott2003,Naulin2005,Madsen2017}. Here, we are interested in how the angular momentum anchors to the background magnetic field in the absence of a symmetry, what components appear
in the complete stress tensor beside the ever present Reynolds stress and the impact of high fluctuation amplitudes and small gradient length scales.

Our derivation rests upon two pillars: (i) a full-F gyro-kinetic formalism, where finite Larmor radius and polarization effects are taken in the \LWL\ and (ii) a \DK ordering of the resulting energy-momentum balance itself.  The \LWL\ is a way to obtain closed expressions in the energy-momentum balance. The main effect of the full-F formalism is the appearance of the density inside flux-surface averages. In order to present the main nonlinearities in a convenient form we introduce the Favre average - a density weighted flux-surface average~\cite{Held2018}. As a natural consequence, the Favre stress emerges, which generalizes the conventional Reynolds stress.
The \DK ordering is necessary to neglect geometric correction terms that would otherwise clutter the resulting expressions and to easily identify fluid moments from velocity space integrals. However, our momentum balance equations 
are valid only up to order three within this ordering.

The magnetic field geometry is arbitrary and we in particular do not invoke a toroidal symmetry. Thus, as long as the orderings hold, our results are applicable to various devices, including tokamaks and stellarators, the reversed field pinch and field-reversed configurations.
Further, we make no assumptions on the form of the gyro-kinetic distribution function and
our results thus apply to gyro-kinetic as well as gyro-fluid models.
At the same time we allow a direct comparison to drift-reduced fluid equations due to the applied \DK ordering.

We carefully recall the definition of angular momentum from the underlying particle Lagrangian in suitable coordinates and construct the inertia tensor with the help of the first fundamental form of general flux-surfaces. This enables us to then construct and discuss a rotational energy balance. Within the energy balance equations we keep terms up to order four in the \DK ordering.

This manuscript is divided into the following parts. In Section~\ref{sec:magnetic} we review the magnetic field representation via flux-coordinates in order to setup suitable poloidal and toroidal angle coordinates.
Our main derivation then proceeds with
 the definition of the gyro-kinetic action in Section~\ref{sec:fundamentals}, which encompasses our assumptions on the model, specifically the \LWL. The \DK ordering scheme is presented in Section~\ref{sec:main}. The latter enables us to then derive
the poloidal and toroidal angular momentum balance up to order three within this ordering and in particular replace gyro-fluid with regular fluid moments in the result. In Section~\ref{sec:momentum_fa} we apply the previously proposed Favre decomposition~\cite{Held2018} in order to identify the signature of relative density 
fluctuations in both known and novel components of the stress tensor.
In Section~\ref{sec:rotational} we derive the relation between angular momentum and angular velocity and identify the inertia tensor. Furthermore, we find the time evolution for the rotational energy
using the previously derived momentum balance. 
Finally, we discuss the significance of our results on various topics discussed in the literature in Section~\ref{sec:discussion}, including the 
electromagnetic field momentum, drift-fluid models, the ion orbit loss mechanism and the transition to simplified geometries.
\ref{app:formulary} provides a formulary intended as a quick reference list of the most often used relations and notations.

\section{Preliminary: the magnetic field in flux-coordinates} \label{sec:magnetic}
A toroidal magnetic field equilibrium can be represented by so-called flux-coordinates $\{\rho, \vartheta, \varphi\}$
(Reference~\cite{Boozer2005} calls them magnetic coordinates)
where the magnetic field lines appear straight 
\begin{align}\label{eq:magtwoform}
    \mathcal B^2 =  \d\psi_p\wedge\d\varphi + \d \psi_t \wedge \d \vartheta
\end{align}
Here, $\psi_p(\rho)$ is the poloidal flux and $\psi_t(\rho)$  is the toroidal flux and we have $\d \psi_p = \iota(\rho)\d\psi_t$ where we introduced the rotational transform $\iota $. Further, $\rho$ is any radially increasing flux label, $\vartheta$ is the poloidal flux angle and $\varphi$ is the toroidal flux angle coordinate. Note that $\vartheta$ increases in the counter-clockwise direction in the poloidal plane while $\varphi$ increases clockwise
if viewed from above to get a right-handed coordinate system.  We emphasize that in general $\varphi$ and $\vartheta$ are different from the geometric angles. In this manuscript we always refer to flux angles when speaking of the toroidal and poloidal angles or directions and will highlight when these angles coincide with the geometric angles. 

There are many different toroidal flux coordinate systems, notably
Hamada and Boozer coordinates \cite{haeseleer,Boozer2005}.
\begin{figure}[htbp]
    \centering
    \includegraphics[width = 0.5 \textwidth]{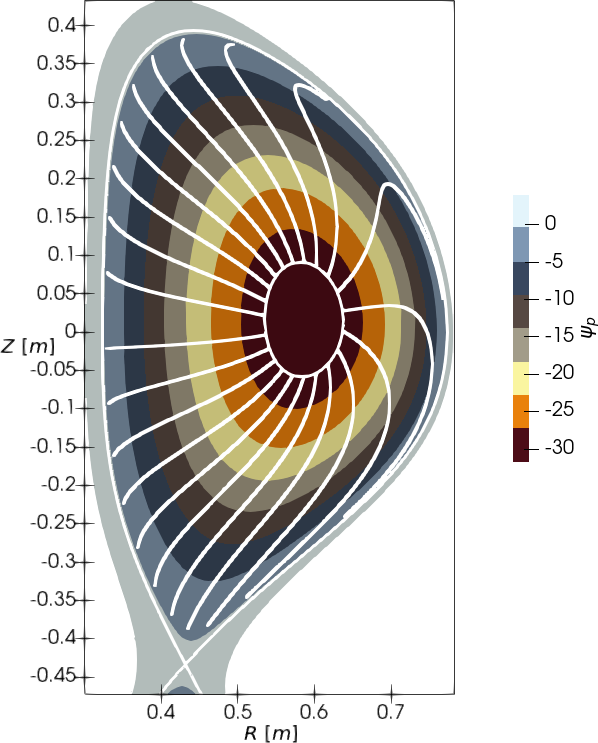}
    \caption{Numerically integrated~\cite{Wiesenberger2017} flux coordinates for an axisymmetric equilibrium. The contours of the poloidal flux label $\psi_p$ are given in colour with white markers for the separatrix and the starting contour for $\vartheta$ integration, while the lines of constant poloidal flux angle $\vartheta$ are given in white as well.}
    \label{fig:flux_grid}
\end{figure}
In Fig.~\ref{fig:flux_grid} we show an example of a numerically integrated~\cite{Wiesenberger2017} flux-coordinate system for an axisymmetric tokamak magnetic field. Here, we show the lines of constant $\psi_p$ in colour and the lines of constant poloidal flux angle $\vartheta$ in white. The toroidal flux angle $\varphi$ coincides with the geometric angle. 

The magnetic field $\mathcal B^2 = \d \mathcal A^1$ can be written as a total differential of the magnetic potential
\begin{align}\label{eq:magpotoneform}
    \mathcal A^1 = \psi_p \d \varphi + \psi_t \d \vartheta
\end{align}
which notably identifies $A_\varphi = \psi_p(\rho)$ and $A_\vartheta = \psi_t(\rho)$. At the same time $\d \mathcal B^2 = \d\circ\d \mathcal A^1 = 0$ immediately as $\d\circ\d =0$ for the exterior derivative $\d$.
This is the coordinate-free expression of vanishing divergence.

We formulate Eqs.~\eqref{eq:magtwoform} and \eqref{eq:magpotoneform} in terms of differential forms,
which we here introduce because the
gyro-kinetic theory heavily relies on them (for an excellent introduction to differential geometry for physicists see Frankel's text~\cite{Frankel}).
An interesting (if somewhat aloof) property of using differential forms is that
they (and therefore the magnetic field) can be defined without the existence of a metric tensor. 
Recall for example that the 1-form $\d\vartheta$ symbolizes the planes that are constructed by
keeping $\vartheta$ constant and varying $\rho$ and $\varphi$, which is a purely topological operation. 
In contrast, the gradient basis vector
$\vn\vartheta$ is the vector that is perpendicular to the planes
of constant $\vartheta$, which requires a metric to define.

We are of course aware of the practicality that the physicist's notation of
Eq.~\eqref{eq:magtwoform} provides
\begin{align}\label{eq:magvector} 
\vec B = {\vn\psi_p \times \vn \varphi}+{\vn\psi_t\times \vn\vartheta}
\end{align}
We are here able to identify the poloidal $\vec B_p:=\vn\psi_p\times\vn\varphi$ and toroidal $\vec B_t:={\vn\psi_t\times \vn\vartheta}$ parts of the magnetic field vector $\vec B$. 
With the choice of signs in Eq.~\eqref{eq:magvector} and assuming $\vn\psi_p$ points radially outwards, we get a left-handed field-line winding when going in the positive $\varphi$ direction since $\vec B_p \sim -\etheta$. 
Furthermore, notice the useful properties
\begin{align} \label{eq:directional_varphi}
    \vn \psi_p = \ephi \times\vec B \\
    \vn \psi_t = \etheta\times \vec B \label{eq:directional_vartheta}
\end{align}
where $\ephi$ and $\etheta$ are the covariant basis
vectors, that is the vectors that generate the directional derivatives along $\varphi$ and $\vartheta$,
or in other words, $\ephi$ is the tangent vector to the line that we get when keeping $\rho$ and $\vartheta$ constant and varying $\varphi$ ($\etheta$ analogous). We emphasize that we mean these two vectors when we speak of toroidal $\ephi$ and poloidal $\etheta$ directions in contrast to the $\vn\varphi$ and $\vn\vartheta$ directions. 
For example, in Fig.~\ref{fig:flux_grid} $\ephi$ points perpendicularly out of the plane while $\etheta$ is tangent to the contours
of $\psi_p$ (!) and in particular does not point in the same direction as $\vn\vartheta$, which has component out of the flux-surface as well.

When we deal with a symmetric field independent of the geometric
toroidal angle, we will choose $\varphi$ as the geometric toroidal
angle and keep $\vartheta$ as a flux-coordinate with $\vn\vartheta\cn\varphi = \vn \rho \cn\varphi = 0$ as we do in Fig.~\ref{fig:flux_grid}. This type of coordinates is
known as symmetry flux or PEST coordinates \cite{Grimm1983}. Notice that we 
do not use the geometric poloidal angle since we want to keep the form Eq.~\eqref{eq:magtwoform}. A useful property of this type of 
coordinate is that $qR^2/\sqrt{g} \equiv I(\rho)$ is a flux function, which allows us to write
\begin{align}\label{eq:symmetric_magnetic_field}
    \vec B = I(\rho) \vn\varphi + \vn\psi_p \times\vn\varphi
\end{align}


Last, note that all flux coordinates are problematic when an X-point with $\vn\psi_p=0$ is present in or close to the domain of interest.
In fact, any coordinate system with a flux label as the first coordinate is problematic when an X-point is present~\cite{Wiesenberger2018}. 
On the one hand the poloidal flux $\psi_p$ is continuous and well-defined across the separatrix. However,
the toroidal flux $\psi_t$ as well as the poloidal flux angle $\vartheta$ 
are only well-defined up to but not including or across the separatrix and furthermore $\iota^{-1}$ diverges on the separatrix.
This is expected since the poloidal component of $\vec B$ vanishes at the X-point. 
In practice, 
the divergence manifests for example in Fig.~\ref{fig:flux_grid} where the
coordinate lines for $\vartheta$ are distorted when
getting close to the separatrix on the low field side of the tokamak.

Last, we introduce the flux surface average (see for example \cite{haeseleer}) as an average over a
small volume - a differential shell centered around the flux-surface.
We define
\begin{align} \label{eq:fsa_vol}
\langle f \rangle (\psi_p) :=& \frac{\partial}{\partial v} \int_\Omega \dV f 
 = \int_{\psi_p}  \frac{f(\vec x)}{|\vn v|} \dA
\end{align}
 where we define $v(\psi_p) := \int_0^{\psi_p} \dV$ as the volume
flux label. In flux coordinates we have $\dA = \sqrt{g} |\vn\rho| \d\vartheta\d\varphi$.
The average fulfills the identity
\begin{align}
\RA{ \nc \vec j} &= \frac{\partial}{
  \partial v} \RA{\vec j \cdot \vn v} 
  \label{eq:fsa_identities}
\end{align}
Also note that for any divergence free vector field $\nc \vec j = 0$ and a flux function $f(\psi_p)$ we have
\begin{align}\label{eq:fsa_zero_div}
    \langle \nc( \vec j f)\rangle = 0
\end{align}
which is proven straightforwardly.

In summary, using flux coordinates for the following derivation defines suitable angle coordinates as well as poloidal and toroidal directions. We expect the resulting expressions to be valid for any flux coordinate system within the closed field-line region up to the separatrix. We remark that the numerical issues of flux coordinates close to the separatrix do not affect the theoretical results presented here.

\section{Fundamentals of Hamiltonian dynamics} \label{sec:fundamentals}

\subsection{Model definition} \label{sec:model}
In this section we define our gyro-kinetic model and discuss the approximations that
go into it. Our goal is to set up a model suitable for edge and scrape-off layer conditions. Literature on the derivation of gyro-kinetic models based on Lie-transform perturbation theory
include the rather technical review~\cite{BrizardHahm} and references therein. A friendlier tutorial can be found in~\cite{Krommes2012} or the more recent~\cite{Tronko2018}.
Here, we start directly with the gyro-centre Poincar\'e 1-form expressed in the transformed phase-space coordinates $\vec Z:=\{\vec X, w_\parallel, \mu, \theta\}$, with gyro-centre coordinate $\vec X$, parallel canonical moment $w_\parallel$, magnetic moment $\mu$, gyro-angle $\theta$
\begin{align}
\gamma := \left( q \vec A + mw_\parallel  \bhat \right)\cdot{\d \vec X}
+\frac{m}{q}\mu\d\theta
\label{eq:particle_oneform}
\end{align}
with species mass $m$ and charge $q$
and we omit the species label. We have the magnetic background potential $\vec A\cdot \d \vec X \equiv \mathcal A^1$ from Eq.~\eqref{eq:magpotoneform} and the background magnetic field unit vector $\bhat := \vec B/B$.
In flux-coordinates Eq.~\eqref{eq:particle_oneform} explicitly reads 
\begin{align}\label{eq:spatial_oneform}
\gamma = (q\psi_t + mw_\parallel b_\vartheta )\d \vartheta+ (q\psi_p + mw_\parallel b_\varphi) \d \varphi +\frac{m}{q}\mu\d\theta
\end{align}
We remark that this 1-form is already enlightening because it immediately identifies 
\begin{align}\label{eq:toroidal_canonical_momentum}
\gamma_\varphi = q\psi_p + mw_\parallel b_\varphi
\end{align}
as the toroidal
angular momentum and 
\begin{align}\label{eq:poloidal_canonical_momentum}
\gamma_\vartheta = q\psi_t + mw_\parallel b_\vartheta
\end{align}
 as the poloidal angular momentum. Recall here that angular momentum is defined as the canonically conjugate momentum to the angle coordinate. In anticipation of the following discussion we here remark that $q\psi_p$ and $q\psi_t$ will lead to the toroidal and poloidal components of the \ExB velocity contribution. The parallel velocity contribution is given by the two components of the magnetic field unit vector $b_\vartheta$ and $b_\varphi$ as expected. Unfortunately however, the definitions for toroidal and poloidal angular momentum in Eqs.~\eqref{eq:toroidal_canonical_momentum} and \eqref{eq:poloidal_canonical_momentum} are \textit{not} coordinate invariant and therefore care must be taken when comparing results from different coordinate systems.
 This is evident since the value of $b_\varphi$ and $b_\vartheta$ depend on the choice of coordinates. Physically, we attribute this to different reference points/axes for the rotation that different angle coordinates entail.

The symplectic 2-form, defined by the Poincar\'e 1-form,
$w:=\d\gamma$, defines the geometry of phase-space much the same way the metric tensor $g$ defines the geometry of ordinary space. The difference is
that $\omega$ defines areas instead of distances and is skew-symmetric instead of symmetric (see \cite{Frankel}).
In 6-dimensional phase-space coordinates we have
\begin{align}
\omega_{ij} &= \frac{\partial \gamma_j}{\partial Z^i} - \frac{\partial \gamma_i}{\partial Z^j}
\nonumber\\ \omega &=
\begin{pmatrix}
-q (\vec B^*\times)
& -m \bhat
& 0 & 0 \\
m\bhat^\T
& 0 & 0 & 0  \\
0 & 0 & 0 & \frac{m}{q} \\[6pt]
0 & 0 & -\frac{m}{q} & 0
\end{pmatrix}
\label{eq:omega}
\\
q\vec B^* &:= q\vec B + m w_\parallel\vn\times \bhat\\
B^*_\parallel &= \vec B^*\cdot\bhat = B + \frac{m w_\parallel}{q}(\vn\times \bhat)_\parallel
\end{align}
Note the covariant vector components 
$b_i$ (with $\bhat^\T := (b_1, b_2, b_3)$) and
the appearance of the determinant of the metric tensor $g$ in the definition of 
the cross-product $(\vec B^*\times)_{ij} := \sqrt{g} \eps_{ikj} B^{*k}$ with contravariant components $B^{*k}$. 

The phase space volume $\vol := \omega\wedge\omega\wedge\omega = \sqrt{\det(\omega)}\d^6Z$ reads
\begin{align}
\sqrt{\det( \omega)} \d^3X\d w_\parallel\d \mu \d\theta &= m^2\sqrt{g}|B_\parallel^*| \d^3X\d w_\parallel\d \mu\d \theta
\label{eq:volume_form}
\end{align}
Notice that the volume form is proportional to $|B^*_\parallel|$ not just $B^*_\parallel$ as often noted since it needs to remain positive. 
More importantly, it is apparent that the coordinate system possesses a (coordinate) singularity
at $mw_\parallel = - qB/(\vn\times\bhat)_\parallel$, where $B_\parallel^*=0$ and thus $\det(\omega) = 0$. This destroys the symplecticity of the 2-form $\omega$, the volume form
Eq~\eqref{eq:volume_form} vanishes and the inverse of $\omega$ diverges (and thus the equations of motion).
It is
questionable how we can deal with this singularity especially when we later integrate
over the phase-space volume to form the field
equations. Furthermore, when deriving gyro-fluid models terms 
$\propto(B_\parallel^*)^{-1}$ prevent identifying velocity
space moments that involve $B_\parallel^*$ in the volume element. This problem is often ignored in the literature or
circumvented by requiring
$(\vn\times \bhat)_\parallel =0$ and we will follow this approach
in this work. For a low-$\beta$ stellarator $\vn\times \vec B=0$, however for general tokamak magnetic fields the requirement is only approximately fulfilled. 
As we will show in Section~\ref{sec:simplified-geometries} the problem is also resolved by simplifying the magnetic field to purely toroidal or poloidal. 
Interestingly, the requirement $(\vn\times\bhat)_\parallel = 0$ relates to the integrability condition for vector fields perpendicular to the magnetic field. The Frobenius theorem~\cite{Frankel} 
states that
planes perpendicular to $\vec \bhat(\vec x)$
everywhere exist in the sense
that there exist functions $\lambda(\vec x)$ and $f(\vec x)$ such that
$\lambda(\vec x)\bhat(\vec x) = \vn f$ if and only if $(\vn\times\vec \bhat)\cdot \bhat = 0$.
In other words we surmise that the existence of drift-planes is a prerequisite
for gyro-kinetic and -fluid models. 

Our Hamiltonian reads
\begin{align}\label{eq:hamiltonian}
H &:= \frac{\left(mw_\parallel- q\mathcal A_{1,\parallel}\right)^2}{2m}   + \mu B + q\Psi \nonumber \\
&\equiv \frac{1}{2}mw_\parallel^2 + \mu B + H_f 
\end{align}
with the effective gyro-centre potentials 
\begin{align}
q\mathcal A_{1,\parallel} &:= qA_{1,\parallel} + \frac{m\mu}{2q B}\Delta_\perp A_{1,\parallel} \\
q\Psi&:= q \phi + \frac{m\mu}{2q B}\Delta_\perp \phi - \frac{1}{2} m \left(\frac{\np \phi}{B}\right)^2 \label{eq:generalized_potential_phi}
\end{align}
where we define the field Hamiltonian $H_f := q\Psi - qw_\parallel \mathcal A_{1,\parallel} + q^2 \mathcal A_{1,\parallel}^2 /2m$ to contain all terms dependent on
the electromagnetic field perturbations $\phi$ and $A_{1,\parallel}$.
The potential $\phi$ is in fact a first order term where the zeroth order $\phi_0$ has been neglected.
The first order perturbation $\mathcal A_{1,\parallel}$ is not to be confused with the zeroth order magnetic field potential $\mathcal A^1$. Finally, see Table~\ref{tab:operators} in \ref{app:formulary} for definitions of $\np$ and $\Delta_\perp$.
Here, we follow~\cite{Scott2010,Brizard2011} and use the Hamiltonian formulation
with $mw_\parallel:= mv_\parallel + q\mathcal A_{1,\parallel}$ such that
the electromagnetic field variations appear in the Hamiltonian only
and do not disturb the symplectic geometry~\eqref{eq:particle_oneform}.
We note that we
\begin{enumerate}
    \item neglect all terms $k_\perp^3\rho_0^3$ with gyro-radius $ \rho_0 := \sqrt{2B\mu m}/eB$ and higher in the Hamiltonian (this especially neglects the second order guiding centre contributions, which according to \cite{Brizard2013} leads to guiding centre drifts in the polarization equation). In particular, both the polarization contribution (the last term in Eq.~\eqref{eq:generalized_potential_phi}) as well as the finite Larmor radius effects are taken in the \LWL~\cite{Held2020}.
    \item neglect compressional Alfv\'en waves entering through $\vec{A_{1,\perp}}$~\cite{Hahm2009}
    \item neglect all terms non-linear in the magnetic potential $A_{1,\parallel}$ (except in the parallel kinetic energy). This approximation implies the absence of $A_{1,\parallel}$ terms in the polarization and of $\phi$ terms in the parallel Amp\`ere law~\cite{Hahm2009}
equation and vice versa $\phi$ terms in the parallel Amp\`ere law and therefore decouples the two equations, which is numerically desirable~\footnote{Desirable might be an understatement. We are not aware of any successful attempts to numerically solve the completely coupled set of equations in a turbulence simulation.}
\end{enumerate}
Our model is comparable to Reference~\cite{Madsen2013} with the difference that we additionally take the \LWL\ in the gyro-average operator.
We note that with our approximations the Hamiltonian formulation with $w_\parallel$ is entirely equivalent to the symplectic formulation using $v_\parallel$ in the sense that the resulting equations are the same. The Hamiltonian formulation is more convenient here since $\gamma$ is time-independent. We also remark that the gyro-average and polarization corrections in our gyro-kinetic model Eq.~\eqref{eq:hamiltonian} resemble the second order guiding centre transformation terms in guiding-centre models~\cite{Madsen2010,Jorge2017}.
However, since we logically start with and approximate a gyro-kinetic model we will keep referring to our model as gyro-kinetic. 

We introduce the gyro-kinetic particle
distribution function $F(\vec Z, t)\equiv F(\vec X, w_\parallel, \mu, t) $ (independent of gyro-angle $\theta$, which is averaged out). The
Vlasov equation states
\begin{align}\label{eq:vlasov}
    \frac{\d}{\d t } F = \frac{\partial F}{\partial t} + {\dot Z^i} \frac{\partial F}{\partial Z^i} = S
\end{align}
Here, $S$ is a general kinetic source term in gyro-centre phase-space $S(\vec X, w_\parallel, \mu, t)$. With the kinetic source function $S$ we formally represent effects like for example non-elastic collisions, plasma-neutral interactions, heating of the plasma, or plasma fuelling and bear in mind that detailed expressions for $S$ are not part of this manuscript. We call $S$ a source understanding that it can act as a sink as well.


Next, with the 1-form $\gamma$ in Eq.~\eqref{eq:particle_oneform} and the Hamiltonian $H$ in Eq.~\eqref{eq:hamiltonian} we can define a particle Lagrangian
\begin{align}\label{eq:particle_lagrangian}
L_p := \gamma_i \dot Z^i - H
\end{align}
Together with the volume form in Eq.~\eqref{eq:volume_form} and
the phase space distribution function $F$ we can then
define the system Lagrangian
$\mathcal L_p:= \sumsp \int \vol(\vec Z) F(\vec Z, t) L_p(\vec Z, \dot {\vec Z},t)$,
where we sum over species. 
Finally, we close the system
with a field Lagrangian and propose the action integral
\begin{align}\label{eq:action}
\mathcal S :=  \sumsp \int \d t
\int \d V \d w_\parallel \d \mu\d\theta m^2 B F (\gamma_i \dot Z^i -H) \nonumber\\
- \int \d t \int\dV \frac{ (\np A_{1,\parallel})^2}{2\mu_0}
\end{align}
where $\d V := \sqrt{g}\d^3 X$ is
the spatial volume form.
The action in Eq.~\eqref{eq:action} plus the Vlasov equation~\eqref{eq:vlasov}
are the central relations in every gyro-kinetic model. They
completely
define the system that we investigate. In particular this means that
$\mathcal S$ contains all approximations to our model and that the
following calculations are exact. 

We remark that
\begin{enumerate}
    \item the neglect of the electric energy $\vec E^2$ in the field part of Eq.~\eqref{eq:action} leads to quasineutrality (that is a vanishing right hand side in Eq.~\eqref{eq:variation_phi})
    \item the magnetic field energy in \eqref{eq:action} neglects the $A_{1,\parallel}(\vn\times \vec B)_\parallel$ contribution from the background magnetic field. This leads to the omission of the background equilibrium current in the Amp\`ere equation~\eqref{eq:variation_apar}. This approximation is in line with $(\vn\times\bhat)_\parallel=0$.
\end{enumerate}

\subsection{The Vlasov-Maxwell equations}\label{sec:vlasov-maxwell}
In the Lagrangian picture~\cite{Sugama2000} the equations of motion can be
retrieved from the action Eq.~\eqref{eq:action} by expressing $\vec Z = \vec Z ( \vec Z_0, t_0; t)$, using $F(Z,t) = F_0(Z_0,t_0)$ by the Vlasov equation~\eqref{eq:vlasov}, taking the integration to
the initial positions and time\footnote{Technically, here we also need to know that the volume form is conserved in time $B(z)\d^6 Z = B(z_0)\d^6z_0$, something that we will need to show explicitly.} and then varying $\delta \mathcal S/ \delta \vec Z = 0$. This indeed recovers the
Euler-Lagrange equations
\begin{align}\label{eq:euler_lagrange}
    \frac{\d}{\d t}\frac{\partial L_p}{\partial \dot Z^i} - \frac{\partial L_p}{\partial Z^i} = 0
\end{align}
The application of the
Euler Lagrange equations~\eqref{eq:euler_lagrange} yields the Hamilton equations of motion
 (using $\d \gamma_i/\d t = \dot Z^j \partial \gamma_i /\partial Z^j$)
 \begin{align}\label{eq:hamilton}
 Z_H^i \omega_{ij} = -\partial_j H \ \leftrightarrow\ i_Z\omega = -\d H 
\end{align}
where we define $Z_H^i \equiv \dot Z^i$ as the components of the 
Hamiltonian vector field on phase space. Here, $i_Z$ is the inner product with the vector field $Z_H$ and $\d$ is the total differential.
The particle trajectories are given by the streamlines of $Z_H$ (with $J := \omega^{-1}$)
\begin{align}\label{eq:inverse_hamilton}
    \frac{\d Z^i}{\d t} = Z_H^i = J^{ij}\frac{\partial H}{\partial Z^j}
\end{align}
The time-derivative of any phase-space function along the trajectory is then given by
\begin{align}
    \frac{\d}{\d t} f(\vec Z, t) = \frac{\partial f}{\partial t} + \dot Z^i \frac{\partial f}{\partial Z^i} =  \frac{\partial f}{\partial t} + Z_H^i \frac{\partial f}{\partial Z^i}
\end{align}
Here and in the following we use $\dot Z$ synonymously with $Z_H$.
In particular, the derivative of the Hamiltonian gives
\begin{align}\label{eq:Hamilton_Dot}
    \frac{\d}{\d t} H(\vec Z, t) = \frac{\partial H}{\partial t} + Z_H^i \frac{\partial H}{\partial Z^i} = \frac{\partial H}{\partial t}
\end{align}
where we use Eq.~\eqref{eq:inverse_hamilton} and the antisymmetry of $J$. 

Explicit expressions for the inverse of the symplectic 2-form Eq.~\eqref{eq:omega} and the gradient
of the Hamiltonian~\eqref{eq:hamiltonian} are  
\begin{align}
J &=
\begin{pmatrix}
\frac{1}{qB} (\bhat\times) & \frac{1}{mB}\vec B^* & 0 & 0 \\[6pt] 
-\frac{1}{mB} \vec B^{*\T} & 0 & 0 & 0  \\[6pt]
0 & 0 & 0 & -\frac{q}{m} \\[6pt]
0 & 0 & \frac{q}{m} & 0
\end{pmatrix}
\label{eq:Pi}
\\
(\partial H )^\T&=
\begin{pmatrix}
 \mu B \vn\ln B + \vn H_f & m v_\parallel  & B^\# & 0 
\end{pmatrix}
\label{eq:nablaH}
\end{align}
with $(\bhat\times)^{ij} := \sqrt{g}^{-1} \epsilon^{ikj}b_k$ and $\vn H_f = - v_\parallel q\vn \mathcal A_{1,\parallel} + q\vn\Psi$. The
$\mu$ component of $\partial H$ contains corrections due to the
fluctuating electric field $B^\# := B + m\Delta_\perp \phi /2q^2B$.
An explicit expression for the components of $Z_H$ (or $\dot Z$) can now be formed given Eqs.~\eqref{eq:Pi} and \eqref{eq:nablaH} (with
$mv_\parallel \equiv mw_\parallel - q\mathcal A_{1,\parallel}$) 
\begin{align}
\label{eq:xdot}
\dot{\vec X} =& \frac{1}{B}\left( \vec B^* v_\parallel +\frac{1}{q}\bhat\times\vn H\right)
\nonumber\\
=&
\frac{1}{B} \left(
\vec B v_\parallel + \frac{mv_\parallel^2}{q} \vn\times\bhat  + \frac{\mu B}{q}\bhat\times \vn \ln B 
\right.\nonumber\\ 
&\ +\left. v_\parallel \vn\times \mathcal A_{1,\parallel}\bhat + \bhat\times \vn\Psi\right)
\\
\label{eq:wpardot}
m\dot w_\parallel =& -\frac{\vec B^*}{B}\cn H
\nonumber\\
=& -\frac{1}{B}\left(
\vec B + \frac{mv_\parallel}{q} \vn\times\bhat + \vn\times \mathcal A_{1,\parallel}\bhat\right) 
\nonumber\\ &\quad
\cdot\left(\mu B\vn\ln B + q\vn\Psi \right)
+ q\dot{\vec X}\cdot\vn \mathcal A_{1,\parallel}
\\
\label{eq:mudot}
\dot \mu =& 0 \\
\label{eq:thetadot}
\dot\theta =& \frac{qB}{m} + \frac{\Delta_\perp\phi}{2B}
\end{align}
The phase space volume $\vol = \omega\wedge\omega\wedge\omega$ is conserved along the particle trajectories
\begin{align}
\frac{\d}{\d t}\vol = \mathcal L_Z \vol = (\d i_Z \omega)\wedge\omega\wedge\omega = 0
\label{}
\end{align}
where the Lie derivative on differential forms
is given by Cartan's formula~\cite{Frankel}
$\mathcal L_Z \alpha = \d (i_Z \alpha) + i_Z\d\alpha$ and
per definition $\d i_Z\omega = -\d\circ\d H = 0$.
In coordinates $\d(  i_Z \vol) = 0$ reads
\begin{align} \label{eq:cons_volume}
    \frac{1}{\sqrt{\det(\omega)}}\partial_i \left(\sqrt{\det(\omega)} Z_H^i\right) =0
\end{align}
The conservation of volume thus translates to a vanishing divergence of the Hamiltonian vector field
in phase space
\begin{align}
\nc ( B\vec{ \dot X} ) + \frac{\partial}{\partial w_\parallel} (B\dot w_\parallel) = 0
\label{eq:liouville}
\end{align}
Notice that volume conservation does not mean that $B$ is conserved
along particle trajectories, we rather have $\dot B = \vec {\dot X}\cn B$.

The conservation of the particle distribution function $F(\vec X, v_\parallel, \mu, t)$ is expressed by the gyro-kinetic Vlasov equation $\d F/\d t = S$,
which together with phase space volume conservation~\eqref{eq:liouville} reads in conservative form
\begin{align}\label{eq:vlasov_conservative}
\frac{\partial \left(BF\right)}{\partial t} +
\nc\left( BF\dot{\vec X}\right) + \frac{\partial \left(BF\dot w_\parallel\right) }{\partial w_\parallel} = BS
\end{align}
The Vlasov-equation~Eq.~\eqref{eq:vlasov_conservative} together with the equations of motion Eq.~\eqref{eq:xdot}-\eqref{eq:mudot} forms the first half of the Vlasov-Maxwell system.

In order to derive the Maxwell equations we first  define the velocity space moment operator~\cite{brizard92}
\begin{align}
\|\zeta\| :=\int \d w_\parallel \d \mu \d\theta m^2 B F\zeta
\label{eq:fluid_moments}
\end{align}
where $\zeta(\vec X,w_\parallel, \mu,t)$ is any function defined on phase-space and the integration encompasses the entire velocity space. Notice that we name the first
few fluid moments $N:=\|1\|$, $NW_\parallel := \|w_\parallel\|$ and $P_\perp := \|\mu B\|$ and give a comprehensive list in~\ref{sec:fluid_moments}.

We also define the moment operator for the source function $S$ analogous to the  velocity space moment operator for the gyro-kinetic distribution function $F$ in Eq.~\eqref{eq:fluid_moments}
\begin{align}\label{eq:source_moments}
    \|\zeta\|_S := \int \d w_\parallel \d \mu\d \theta m^2 B S\zeta
\end{align}
Analogous to the moments of $F$ we name the source moments $S_N:=\| 1\|_S$, $S_{P_\perp} = \| \mu B\|_S$, etc.

Using Eq.~\eqref{eq:vlasov_conservative} together with the fact that
$\partial/\partial t$ and $\vn$ commute with the velocity integral and
$F$ vanishes for $w_\parallel = \pm \infty$ 
we find the important identity~\cite{brizard92}
\begin{align}
\frac{\partial}{\partial t}\|\zeta\| + \nc
\|\zeta\vec{ \dot X}\|
= \bigg\|\frac{\d\zeta}{\d t}\bigg\| +\| \zeta \|_S
\label{eq:zeta_general}
\end{align}

The
variation of the action~\eqref{eq:action} with respect to $\phi(\vec x)$ yields
the quasi-neutrality equation
\begin{align}\label{eq:variation_phi}
    \frac{\delta \mathcal S}{\delta \phi} = \frac{\delta}{\delta \phi(\vec x)} \sumsp \int \d V \d w_\parallel \d \mu\d\theta m^2 BF H  = 0
\end{align}
and with respect to $A_{1,\parallel}$ the parallel Amp\`ere law
\begin{align}\label{eq:variation_apar}
    \frac{\delta \mathcal S}{\delta A_{1,\parallel}} = \frac{\delta}{\delta A_{1,\parallel}(\vec x)} \sumsp \int \d V \d w_\parallel \d \mu \d\theta m^2 BF H
    \nonumber\\
    +  \frac{\delta}{\delta A_{1,\parallel}(\vec x)} \int \dV \frac{ (\np A_{1,\parallel})^2}{2\mu_0} = 0
\end{align}
where we used that $\gamma$ does not depend on either $\phi$ or $A_\parallel$.
Now, recall the variational derivative. For each $\zeta\in \{\phi, A_{1,\parallel}\}$
and $H= H(\zeta, \np\zeta, \Delta_\perp \zeta)$ 
we have
 \begin{align}\label{eq:variational_derivative}
      \frac{\delta}{\delta\zeta(\vec x)} \int_{R^3}\d^3x' F B H=  F B\frac {\partial H}{\partial\zeta}  \nonumber\\
     -\nc\left( h F B\frac {\partial H}{\partial\np\zeta}\right) +
    \Delta_\perp\left(F B\frac {\partial H}{\partial\Delta_\perp\zeta}\right) 
 \end{align}
Notice the appearance of $FB$ inside the divergence/Laplace
operators.
Carrying out the variations with the help of Eq.~\eqref{eq:variational_derivative} in the polarization and Amp\`ere equations~\eqref{eq:variation_phi} and \eqref{eq:variation_apar} and identifying the
velocity space moments~\eqref{eq:fluid_moments} yields\footnote{The attentive reader will notice that Eqs.~\eqref{eq:polarization_fluid} and
\eqref{eq:induction_fluid} are
only semi-elliptic since the projection tensor
$h$ is only positive semi-definite. Concerns about existence and uniqueness of solutions are dealt with under "degenerate partial differential equations" in the mathematical literature.
In particular, the field of stochastic differential equations contains a solution to the Dirichlet problem, see for example Reference~\cite{Oeksendal}.
}
\begin{align}
    \sumsp  qN - \nc\vPgy  &= 0 \label{eq:polarization_fluid}\\
    \sumsp q NU_\parallel + \nc( \vMgy\times\bhat) &= -\frac{1}{\mu_0}\Delta_\perp A_\parallel \label{eq:induction_fluid}
\end{align}
with $mU_\parallel\equiv \left( mW_\parallel - q\mathcal A_{1,\parallel}\right)$, $ j_{\mathrm{mag},\parallel} :=  \nc( \vMgy_\perp\times\bhat) = (\vn\times\vMgy)\cdot\bhat - (\vn\times\bhat)\cdot\vMgy$
and the gyro-kinetic polarization and magnetization densities
\begin{align}
    \vPgy  &:= -\sumsp \left[\np\left(\frac{mP_\perp }{2qB^2}\right) +\frac{ m N \np \phi}{B^2}\right] \label{eq:gyro_polarization}\\
    \vMgy_\perp &:= \sumsp\bhat \times\vn \left(\frac{m(Q_\parallel + U_\parallel P_\perp)}{2qB^2}\right)
\end{align}
Note that the parallel component of the polarization current $\vec j_{\mathrm{pol}}\cdot\bhat = \partial \vPgy\cdot \bhat/\partial t = 0$ vanishes in Eq.~\eqref{eq:induction_fluid}.
Also, the parallel part of the magnetization density $\vMgy_\parallel:= -\|\mu\| \bhat\equiv -P_\perp \bhat /B$ does not contribute to the parallel magnetization current.


In total, we now explicitly derived the equations of the Vlasov-Maxwell
system in Eq.~\eqref{eq:vlasov_conservative},\eqref{eq:polarization_fluid} and
\eqref{eq:induction_fluid} together with the equations of motion in
\eqref{eq:xdot}-\eqref{eq:mudot}.

\subsection{Interlude: relation between gyro-fluid and fluid moments} \label{sec:gyro-fluid}
Gyro-fluid quantities like $\|1\| = N(\vec X, t) $ or $\|v_\parallel\| = U_\parallel(\vec X, t)$ are given in gyro-centre coordinates $\vec X$
and are thus not directly comparable to the physical fluid quantities, which we denote with lower case letters $n(\vec x,t):= \int\d^3v f(\vec x, \vec v, t)$, $u_\parallel(\vec x,t) := \int \d^3 v v_\parallel f(\vec x,\vec v, t)$ ..., where $f(\vec x,\vec v,t)$ is the distribution function in
particle phase-space (and we here overburden the use of $v$ as the velocity on top of the volume flux-label).
We need to use the gyro-kinetic phase-space coordinate transformations to transform between particle and gyro-kinetic phase-space moments. Helpfully, Reference~\cite{BrizardHahm} relates the coordinate transformation to the variational derivative of the action.
With our action~\eqref{eq:action} we obtain
\begin{align}
    ||\xi||_{\vec{v}}= \|\zeta\| + \Delta_\perp \left( \frac{m\|\mu B\zeta\|}{2qB^2}\right) + \nc\left( \frac{m \|\zeta\| \vn_\perp \phi}{B^2}\right)
    \label{eq:gyro-fluid-trafo}
\end{align}
where $\xi$ is $\zeta$ transformed to particle coordinates and $\|\xi\|_\vec{v} := \int \d^3v \xi f $ is the particle phase-space moment operator.  Thus, $||\xi||_\vec{v}$ is the physical fluid moment corresponding to $\|\zeta\|$. In Eq.~\eqref{eq:gyro-fluid-trafo} we immediately see that the actual fluid moment $\|\xi\|_\vec{v}$
equals the gyro-fluid moment $\|\zeta\|$ up to an order $\mathcal O(\rho_0^2 k_\perp^2)$ correction. For example the density transforms as
\begin{align}
    n = N + \Delta_\perp \left(\frac{ mP_\perp}{2q^2B^2}\right) + \nc \left( \frac{ mN }{qB^2} \np \phi\right)
\end{align}
The right hand side terms appear exactly in the polarization equation~\eqref{eq:polarization_fluid}, which we obtained from the variational principle. This shows that Eq.~\eqref{eq:polarization_fluid} is the gyrofluid version of quasineutrality $\sumsp qn = 0$.

It is possible to invert the relation between gyro-fluid and fluid quantities. We follow~\cite{Madsen2015,Held2016} and explicitly express
the first two gyro-fluid quantities $N$ and $U_\parallel$ in terms of the true fluid quantities $n$ and $u_\parallel$ in the \LWL\ up to order $(\rho_0k_\perp)^2$.
\begin{align} \label{eq:particle_density}
    N &= n - \Delta_\perp \left(\frac{ mnt_\perp}{2q^2B^2}\right) - \nc \left( \frac{ mn }{qB^2} \np \phi\right)
    \\
    NU_\parallel &= nu_\parallel - \Delta_\perp \left( \frac{m( q_\parallel + u_\parallel p_\perp)}{2q^2B^2}\right)\label{eq:uparallel}
\end{align}
Note that we neglect the potential part in Eq.~\eqref{eq:uparallel} since we miss the corresponding term in the Hamiltonian.

The moments of $S$ transform back to particle phase space analogous to Eq.~\eqref{eq:gyro-fluid-trafo}. This is because the coordinate transformation works for any phase-space function, not just the distribution function $F$.  For example, we have
\begin{align}
    S_{n} = S_{N} + \Delta_\perp \left( \frac{m  S_{P}}{2q^2B^2}\right) + \nc\left( \frac{m S_{N} }{qB^2}\vn_\perp \phi\right)
    \label{eq:particle_source}
\end{align}
where $S_{n}$ is the true fluid particle source term.
We are now able to formulate the only constraint we have for the source term namely that it should conserve the total electric charge via
\begin{align}\label{eq:quasineutral_sources}
    \sumsp q S_n = \sumsp q S_N - \nc\vec S_P= 0
\end{align}
where we define the polarization source
\begin{align}\label{eq:polarization_sources}
    \vec S_P = -\sumsp \left[\np\left(\frac{mS_{P_\perp }}{2qB^2}\right) +\frac{ m S_N \np \phi}{B^2}\right]
\end{align}
Note that Eq.~\eqref{eq:quasineutral_sources} is completely analogous to Eq.~\eqref{eq:polarization_fluid}.


\section{The poloidal, toroidal and parallel momentum balance}\label{sec:main}
\subsection{Poloidal and toroidal \texorpdfstring{\ExB}{TEXT} momentum} 
With the model developed in Sections~\ref{sec:magnetic} and \ref{sec:fundamentals} we are now ready to start
the derivation of the balance equations for the angular momentum density. Keep in mind that we do not assume a toroidal symmetry here. This prohibits us from using Noether's theorem to derive an exact angular momentum balance from the action Eq.~\eqref{eq:action}\cite{Scott2010, Brizard2011}. Instead, we begin by computing the time derivative of $qA_\varphi = q\psi_p$, which is the first part of the toroidal angular momentum~\eqref{eq:toroidal_canonical_momentum}
\begin{align}\label{eq:psipdot}
    q\frac{\d \psi_p}{\d t} =&
    q\vec {\dot X}\cdot \vn \psi_p 
    \nonumber\\
   =& \left(\mu B \frac{\bhat\times\vn\ln B}{B}  + mv_\parallel^2 \frac{\bhat\times\vec \kappa}{B} \right)\cdot \vn \psi_p
    \nonumber\\
    &+ qA_{1,\parallel} v_\parallel \vec K_\kappa \cdot\vn\psi_p
    - \frac{\bhat\times\vn\psi_p}{B}\cdot\vn H_f 
\end{align}
where we separated the field Hamiltonian $H_f$.
Now, to simplify the right hand side of Eq.~\eqref{eq:psipdot} we need to relate
 the variational derivative to ordinary derivatives. Consider a generic Hamiltonian dependence $H(\zeta,\np \zeta, \Delta_\perp\zeta)$ and $\vec \eta := \bhat \times \vn\psi_p/B$
 \begin{align*}
     \vec\eta\cn H = \sum_{\zeta\in \{\phi, A_{1,\parallel}\}} \left\{\frac{\partial H }{\partial \zeta}\vec\eta\cn \zeta
     +\frac{\partial H}{\partial \np\zeta}\cdot \vec\eta\cn\np\zeta \right.\nonumber\\
     \left.+\frac{\partial H}{\partial\Delta_\perp\zeta}\vec\eta\cn\Delta_\perp\zeta\right\}
 \end{align*}
 In order to proceed we need to commute $\vec\eta\cn $ with $\np$ and $\Delta_\perp$. To avoid tedious geometrical correction terms we now introduce a \DK 
ordering~\cite{Simakov2003,Madsen2017}, where we order
\begin{enumerate}
    \item the frequency of turbulent fluctuations compared to the ion gyro-frequency as small
    $\omega/\Omega_i\sim \delta^2\ll 1$, where $\Omega_i = eB/m_i$ 
    \item the derivatives $\nabla_k$ of the dynamical fields as $ \rho_i |\nabla_k \ln F| \sim\rho_i |\nabla_k \ln \phi| \sim \rho_i |\nabla_k \ln A_{1,\parallel}| \sim \rho_i k_\perp \sim \delta$ with ion thermal gyro-radius $\rho_i = \sqrt{m_iT_i}/q_iB$. This in particular orders the \ExB velocity compared to the ion thermal velocity as $u_E /c_{s,i} \sim \delta$ where $c_{s,i} = \sqrt{T_i/m_i}$
    \item all derivatives on the magnetic field (vectors) as $L_B^{-1} \sim |\vn \ln B|\sim 1/R$, where $R$ is the major radius and take $\rho_i/L_B\sim \delta^3$. 
    \item $\rho_i |\npar \ln \phi| \sim\rho_i |\npar \ln A_{1,\parallel}| \sim\rho_i |\npar \ln F|\sim \rho_i k_\parallel \sim \delta^3$ that is parallel derivatives on the magnetic field variation scale. This implies $k_\parallel/k_\perp \sim \delta^2$ 
\end{enumerate}
Note that Reference~\cite{Tronko2018} orders $\rho_i/L_B \sim \delta^4$. However,
this would completely neglect all curvature terms
in our scheme.
In our ordering the Hamiltonian $H_f$~\eqref{eq:hamiltonian} appears to be second order.

We now
neglect all terms of order $\delta^4$ on the right hand side of 
Eq.~\eqref{eq:psipdot}.
With the above orderings we directly have that
$\eta^i\vn_i h^{kl} \sim \delta^3$ and $h^{kl}\vn_l \eta^i \sim \delta^3$. With this and $\partial H/\partial \np \phi = m \np \phi / B^2$ we can order
$m\np \phi\cdot  \eta^i\vn_i \np \phi = m \np \phi \cdot \np (\eta^i \vn_i \phi) + \mathcal O(\delta^5)$.
With similar arguments we can order $\eta^i\vn_i \Delta_\perp \phi = \Delta_\perp (\eta^i\vn_i \phi) + \mathcal O (\delta^5)$.
 Then we have~\cite{Scott2010}
\begin{align}\label{eq:derivative_as_variation}
    F\vec\eta\cn H& = 
     \sum_\zeta\left\{\frac{\delta}{\delta\zeta(\vec x)}\left( \int_{R^3}\d^3x' F  H \right)\vec\eta\cn \zeta
     \right.\nonumber\\
     +\nc&\left.\left[F\frac{\partial H}{\partial \np\zeta}\vec\eta\cn\zeta
     +\np \left(\vec\eta\cn\zeta F\frac{\partial H}{\partial \Delta_\perp\zeta}\right)
    \right. \right. \nonumber\\& \left.\left.
     - 2\np\left(F\frac{\partial H}{\partial \Delta_\perp\zeta}\right)\vec\eta\cn \zeta
     \right]\right\}
\end{align}
This equation is a useful identity and in fact holds for any vector field that commutes with $\np$ and $\Delta_\perp$. It links the ordinary derivative on $H$
 to the variational derivative and correction terms
 that appear as exact divergences.

Summing over all species, integrating over velocity space and inserting our Hamiltonian from Eq.~\eqref{eq:hamiltonian} we get
\begin{align}\label{eq:etagradH}
    \sumsp\|\vec \eta \cdot \vn H\| =  \frac{1}{\mu_0} \Delta_\perp A_{1,\parallel} \vec \eta\cn A_{1,\parallel}
    \nonumber\\
     +\sumsp\nc\left[ -\frac{ mN\np\phi}{B^2}\vec \eta\cn \phi 
     + \np\left(\frac{m\|\mu B\|}{2qB^2} \vec \eta\cn \phi\right)
     \right.\nonumber\\ \left.
     -\np \left(\frac{m\|\mu B\|}{qB^2}\right) \vec \eta\cn \phi 
     - \np\left(\frac{m\|\mu B v_\parallel\|}{2qB^2}\vec \eta\cn A_{1,\parallel} \right)
     \right.\nonumber\\ \left.
     +\np \left(\frac{m\|\mu Bv_\parallel\|}{qB^2}\right) \vec \eta\cn A_{1,\parallel}
     \right]
\end{align}

Now, we focus on the term $q\dot \psi_p = q\dot{\vec X}\cn\psi_p$ on the left hand side of Eq.~\eqref{eq:psipdot}. First we insert $q$ into the velocity space moment equation~\eqref{eq:zeta_general}. We find $\partial_t \|q\| + \nc\|q \vec{\dot X}\| =  \|q\|_S$. 
Under species summation we see that we can identify the polarization equation~\eqref{eq:polarization_fluid}  $\sumsp qN = \nc \vPgy$ and analogously the quasineutrality for the sources Eq.~\eqref{eq:quasineutral_sources} $\sumsp qS_N = \nc\vec S_P$. The next step is to apply the flux-surface average Eq.~\eqref{eq:fsa_vol} to obtain
$\partial_v \left\{ \partial_t \RA{\vPgy\cn v} + \sumsp\RA{\|q \vec{\dot X}\cn v\|} - \RA{\vec S_P \cn v} \right\} = 0$. After volume integration $\int_0^v \d v$ (the inner integration boundary vanishes) and multiplying with $\d \psi_p/\d v$ we obtain
\begin{align} \label{eq:simplified_psipdot}
\sumsp \RA{ \| q\dot \psi_p\|} = -\frac{\partial}{\partial t} \RA{ \vPgy \cdot \vn \psi_p } + \RA{\vec S_P \cn\psi_p}
\end{align}
which recovers the radial part of the polarization current $\vec j_{\mathrm{pol}} \equiv \partial \vPgy/\partial t$. We stress that Eq.~\eqref{eq:simplified_psipdot} is an important 
identity~\cite{Scott2010}. It links the derivative of the poloidal flux or in fact the first part of the toroidal angular momentum of particles
to the polarization current and sources.

Now, we further investigate the terms appearing from Eq.~\eqref{eq:simplified_psipdot} by explicitly inserting our polarization density~\eqref{eq:gyro_polarization}
$-\frac{\partial}{\partial t} \RA{ \vPgy \cdot \vn \psi_p } = \sumsp\frac{\partial}{\partial t} \left[ mN\np\phi\cn \psi_p /B^2 + m\np(\|\mu B\|/2qB^2)\cn\psi_p \right]$.
The second term can be simplified using the dynamical pressure equation  $\d(\mu B)/\d t = \mu B\vec{\dot X}\cn \ln B$ in Eq.~\eqref{eq:zeta_general} yielding $\frac{\partial }{\partial t} \|\mu B\|  =- \nc \|\mu B \vec {\dot X}\| + \| \mu B \vec {\dot X}\cn \ln B\| + \|\mu B\|_S$.
We get the useful identity 
\begin{align}\label{eq:onehalf}
    &\frac{\partial}{\partial t} \RA{  \np\left(\frac{m\|\mu B\|}{2qB^2}\right)\cn\psi_p}
   -\RA{  \np\left(\frac{m\|\mu B\|_S}{2qB^2}\right)\cn\psi_p} \nonumber\\
        &= \frac{\partial}{\partial v} \RA{
    \vn v \cn \left(\frac{m\|\mu B\|}{2qB^2} \vec\eta \cn \phi -  \frac{m\| \mu B v_\parallel \|}{2qB^2} \vec\eta\cn A_{1,\parallel}\right)}
\end{align}
One key ingredient for Eq.~\eqref{eq:onehalf}
is to use $(\vec a \cdot \vn) \vec b = \vn (\vec a\cdot \vec b) - \vec a\times(\vn\times \vec b) - \vec b \times (\vn\times \vec a) - (\vec b \cdot \vn)\vec a$
to show
$\langle \nc(  \vn\psi_p\cn(  \lambda \vec u_\perp) ) \rangle = \partial_v \langle \vn v \cn (\lambda \vec u_\perp \cdot \vn \psi_p) \rangle + \mathcal O (\delta^4)$
in our ordering.
Now, we add the terms Eq.~\eqref{eq:etagradH} and \eqref{eq:onehalf} and use our ordering to eliminate the magnetic field derivatives to get 
\begin{align*}
&\frac{\partial}{\partial t} \RA{\np\left(\frac{ m\|\mu B\|}{2qB^2}\right)\cn\psi_p} + \RA{\|\nabla_\eta H\|} -\RA{  \np\left(\frac{mS_{P_\perp}}{2qB^2}\right)\cn\psi_p}
\\
= &\frac{\partial}{\partial v} \RA{ \frac{\nabla^{v}A_{1,\parallel}\nabla_\eta A_{1,\parallel}}{\mu_0} -\frac{mN\nabla^{v}\phi\nabla_\eta\phi}{B^2 }} 
\\
+& \frac{\partial}{\partial v} \RA{ \frac{m\nabla^v A_{1,\parallel}\nabla_\eta\|\mu Bv_\parallel\|}{qB^2} - \frac{m\nabla^v \phi \nabla_\eta\|\mu B\|}{qB^2} }
\end{align*}
where we use the abbreviation $\nabla^v = \vn v\cn $ and $\nabla_\eta =\vec\eta\cdot \vn$ and imply species summation to present this intermediate result.
We also used that $\vec\eta$ commutes with $\np$ in our ordering and that the flux-surface average of $\vn \cdot (\vec{\eta} h)$ 
vanishes exactly.

Furthermore, we replace the gyro-centre quantities by their particle analogons, which is possible in our ordering since the correction terms are of higher order (see Eqs.~\eqref{eq:particle_density}). Finally, the term $\|\mathcal A_{1,\parallel} v_\parallel\| \vec K_\kappa \cdot\vn\psi_p$ in Eq.~\eqref{eq:psipdot} vanishes  under species summation and the parallel Amp\`ere law to lowest order.
With the help of Eq.~\eqref{eq:directional_varphi} we then finally arrive at
\begin{align} \label{eq:reynolds_flow}
    &\frac{\partial}{\partial t} \sumsp \RA{ m n u_{E,\varphi}}
   +\frac{\partial}{\partial v} \sumsp\RA{
    mn  u_{E,\varphi} \left(u_E^\tv + u_{D}^\tv\right)}
         \nonumber\\
         &- \frac{\partial}{\partial v} \RA{  B_{1,\perp,\varphi}\left( \frac{1}{\mu_0}   B_{1,\perp}^\tv  - \Mem^{,v}\right)}
\nonumber\\
 &=- \RA{ \left( \vec j_f \times \vec B \right)_\varphi} +\sumsp\RA{m S_nu_{E,\varphi}}
\end{align}
Here, we define the \ExB drift $\vec u_E$, the
grad-B drift $\vec u_{\vn B}$, the diamagnetic drift $\vec u_D$, the curvature drift $\vec u_\kappa$, the first order magnetic fluctuations $\vec B_{1,\perp}$ and  the electromagnetic magnetization density $\Mem$ (different from $\vMgy$ by a factor $2$ and fluid instead of gyro-fluid quantities)
\begin{align}
\vec u_E &:= \frac{\bhat\times\vn \phi}{B}  
&& 
\vec u_{\vn B} := t_\perp \frac{\bhat\times\vn \ln B}{qB}
\\
\vec u_{D} &:=\frac{\bhat\times\vn p_\perp}{qnB}  
&&
\vec u_\kappa := (t_\parallel + mu_\parallel^2) \frac{\bhat\times\vec \kappa}{qB}
 \\
 {\vec B}_{1,\perp} &:= {\vn A_{1,\parallel} \times\bhat}
 &&
\vMem :=\sumsp\frac{m\bhat\times \vn (q_\parallel + p_\perp u_\parallel)}{q B^2}
\end{align}
and $\vec b_{1,\perp} := \vec B_{1,\perp}/B$.
Equation~\eqref{eq:reynolds_flow} describes the evolution of the toroidal \ExB angular momentum density and is the first result of this paper. The second term on the left side is the average over the convective acceleration term with radial velocity $u_E^\tv + u_D^\tv$, the sum of \ExB and diamagnetic velocity. In Section~\ref{sec:momentum_fa} we will show that this term can be split into an advective part and components of the turbulent stress tensor. Note that the appearance of the diamagnetic velocity in the gyro-kinetic momentum balance is a consequence of the pressure gradient in the polarization density~\eqref{eq:gyro_polarization} and thus ultimately a gyro-averaging effect. This contrasts to a drift-fluid model where diamagnetic velocity appears as a fluid-drift. The remaining terms on the left hand side are two stress terms stemming from magnetic fluctuations. On the right hand side the Lorentz force originating from the "free" current $\vec j_f := \sumsp qn(\vec u_\kappa + \vec u_{\vn  B} )$ appears and we obtain a momentum source term proportional to the \ExB velocity and the density source $S_n$. 

Another point we note is that
the poloidal analogue of Eq.~\eqref{eq:reynolds_flow} follows immediately.
Recall Eq.~\eqref{eq:directional_vartheta}
together with $\iota\vn\psi_t = \vn\psi_p$ in flux coordinates. This yields
$u_{E,\vartheta}=\vn\phi\cn\psi_t/B^2  = \iota^{-1}u_{E,\varphi}$ and thus from Eq.~\eqref{eq:reynolds_flow} directly follows the equation for the \textit{poloidal} \ExB angular momentum density
\begin{align} \label{eq:reynolds_flow_vartheta}
    &\frac{\partial}{\partial t} \sumsp \RA{ m n u_{E,\vartheta}}
   +\frac{\partial}{\partial v}\RA{
    mn  u_{E,\vartheta} \left(u_E^\tv + u_{D}^\tv\right)}
         \nonumber\\
         &- \frac{\partial}{\partial v} \RA{  B_{1,\perp,\vartheta}\left( \frac{1}{\mu_0}   B_{1,\perp}^\tv  - \Mem^{,v}\right)}\nonumber\\
    &+ \left[\sumsp \RA{ mnu_{E,\vartheta} \left(u_E^\tv + u_{D}^\tv\right)}
         \right.\nonumber\\
        &-  \left.\RA{  B_{1,\perp,\vartheta}\left( \frac{1}{\mu_0}   B_{1,\perp}^\tv  - \Mem^{,v}\right)}\right] \frac{\partial}{\partial v}\ln \iota 
\nonumber\\
 &=- \RA{ \left( \vec j_f \times\vec B \right)_\vartheta} +\sumsp\RA{m S_nu_{E,\vartheta}}
\end{align}
Equation~\eqref{eq:reynolds_flow_vartheta} exhibits a similar structure as Eq.~\eqref{eq:reynolds_flow}
with the additional appearance of the magnetic shear $\sigma:= \partial_v \ln \iota$~\cite{haeseleer}. Depending on its sign the shear term can both dampen and generate poloidal \ExB angular momentum. However, we 
emphasize that the shear appears as a purely geometrical
correction to the poloidal momentum balance. Physically, Eqs.~\eqref{eq:reynolds_flow_vartheta} and \eqref{eq:reynolds_flow} contain the same information
since the two components of the \ExB drift are related.

Before we continue with the identification 
of the various stress terms in Section~\ref{sec:momentum_fa} and a more detailed interpretation of our results,
we first derive the equations for the remaining angular momentum components, namely the poloidal and toroidal angular momentum components stemming from $u_\parallel$. As it turns out we will get the full parallel momentum balance as a by-product.
Finally, recall that both $u_{E,\varphi} = \vn\phi\cn\psi_p/B^2$ and $u_{E,\vartheta}=\vn\phi\cn\psi_t/B^2$ are related to the radial electric field, 
a fact that will lead to the identification of the electromagnetic field angular momentum density in Section~\ref{sec:emfields}.

\subsection{Parallel (angular) momentum}\label{sec:parallel}
We now turn to the parallel terms in the toroidal canonical momentum $\gamma_\varphi = q\psi_p + mw_\parallel b_\varphi$ as well as the poloidal canonical momentum $\gamma_\vartheta = q\psi_t + mw_\parallel b_\vartheta$.
Repeating the ordering scheme from the previous section one could assume that $\partial u_\parallel /\partial t \sim \delta^2$ and argue that therefore only $\mathcal O(\delta^2)$ terms should be kept in our ordering. However, we note that 
 the ions accelerate very slowly. This 
 is because $n\dot u_{\parallel,i}\sim\npar p_i \sim \delta^3$.
 Note that this requires $\np u_{\parallel,i}$ to
 be small as well.
 In contrast, the electron velocity is mainly determined by parallel
 Ohm's law $nu_{\parallel,e} \sim \eta_\parallel^{-1} (\npar p_e + \npar\phi)\sim\mathcal O(1)$ with $\eta_\parallel$ being the parallel resistivity.
 In order to reflect these considerations we
order (in line with Reference~\cite{Simakov2003})
\begin{align}
\partial_t u_{\parallel,i}/\Omega_{c,i}c_s \sim \delta^3    
\end{align}
This ordering mandates that the terms $\partial_t m_i u_{\parallel,i} \sim \partial_t m_i u_{E,\varphi} \sim \delta^3 $ are similar in size. The parallel ion velocity itself is larger than the \ExB velocity but we order its time derivative smaller by the same factor. In total, we again do not assume toroidal symmetry but we do use the \DK ordering and keep terms up to $\mathcal O(\delta^3$).

We start with (for $\eta\in \{\varphi,\vartheta\}$)
\begin{align}
    m\frac{\d}{\d t} (w_\parallel b_\eta) = mw_\parallel \vec {\dot X} \cdot\vn b_\eta + m\dot w_\parallel b_\eta
\end{align}
With the vector triple product rule applied to $(\bhat\times \vn H)\times\vec B^*$ we see $m\dot w_\parallel \bhat = q(\vec {\dot X}\times \vec B) + mw_\parallel \vec{\dot X}\times (\vn\times \bhat)- \mu B\vn\ln B - \vn H_f$.
Next, we note $\vec {\dot X} \cdot \vn b_\eta  = - (\vec{\dot X}\times (\vn\times \bhat))_\eta+  \dot X^i \partial_\eta b_i$ (Notice that we do not use the covariant derivative here since $b_\eta \equiv \bhat\cdot \ehat_\eta$ is a scalar quantity and $(\vec a \cdot \vn \vec b)_\eta\neq \vec a \cdot\vn b_\eta$; the first is the component of a covariant derivative while the second is the directional derivative of $b_\eta$).
Finally, we have
$q(\vec {\dot X}\times \vec B) = mv_\parallel w_\parallel \vec K_\kappa \times \vec B + \mu B \vec K_{\vn B}\times \vec B +
(\bhat\times\vn H_f) \times \vec B/B$.
We thus have
\begin{align}
    m\frac{\d}{\d t} (w_\parallel &b_\eta) =  mv_\parallel w_\parallel (\vec K_\kappa \times \vec B)_\eta + \mu B (\vec K_{\vn B}\times \vec B)_\eta 
    \nonumber\\
    - &b_\eta \npar H_f - \mu B \partial_\eta (\ln B) + mw_\parallel\dot X^i \partial_\eta b_i
     \\
         m\frac{\d}{\d t} w_\parallel &=  -\mu B\npar \ln B - \npar H_f - \frac{m}{q}w_\parallel \vec K_\kappa  \cdot\vn H
\end{align}
The second identity follows immediately from the equations of
motion~\eqref{eq:wpardot}.
In the \DK ordering (and under species summation to make $qv_\parallel A_{1,\parallel}$ vanish) we
can write $mw_\parallel\dot X^i\partial_\eta b_i = mv^2_\parallel  b^i \partial_\eta b_i + \mathcal O( \delta^4)$,
which we interpret as a generalized curvature contribution. 
With similar arguments as in the previous section we can 
recover the variational derivatives in $\sumsp \|b_\eta \npar H_f\|$ using Eq.~\eqref{eq:derivative_as_variation}. However, the remaining terms are all $\mathcal O(\delta^5)$ and can be safely neglected in our ordering. 
Taking the velocity space moment
we arrive at 
\begin{align}
    m&\frac{\partial}{\partial t} \|w_\parallel\| b_\eta + m\nc( \| w_\parallel\vec{\dot X}\| b_\eta) 
    + \|\mu B \|\partial_\eta \ln B 
    \nonumber\\
    &= \|mv_\parallel^2 (\vec K_\kappa \times \vec B)_\eta + \mu B (\vec K_{\vn B}\times \vec B)_\eta \|
    \nonumber\\
    &+\|mv_\parallel^2\| B^i \partial_\eta b_i /B + m\| w_\parallel \|_S b_\eta
    \\
    m &\frac{\partial}{\partial t} \|w_\parallel\| + m\nc( \|\vec{\dot X} w_\parallel\| ) 
    + \|\mu B\|\npar \ln B\nonumber\\
    &= m\| w_\parallel\|_S  
\end{align}
We note that $m\frac{\partial}{\partial t} \|w_\parallel\| = m\frac{\partial}{\partial t} \|v_\parallel\| +q\frac{\partial}{\partial t} \|\mathcal A_{1,\parallel}\|$
and
\begin{align*}
    \| (mw_\parallel& - q \mathcal A_{1,\parallel})\vec{\dot X}\| - m\| v_\parallel\|_S
    \nonumber\\
  = &
  \|mv_\parallel^2\|\left(\bhat + \frac{\vn\times \mathcal A_{1,\parallel} \bhat}{B}\right)  + m NU_\parallel \frac{\bhat \times \vn\psi}{B}
  \nonumber\\
      &+ \frac{m\|mv_\parallel^3\|}{q} \frac{\vn\times\bhat}{B}
 + \frac{m\|\mu B v_\parallel\|}{q} \frac{\bhat\times\vn\ln B}{B}
\end{align*}
Note here that the curvature terms vanish under the divergence in our ordering.
As a final step we again apply the flux-surface average and note
that with implied species summation the term $\partial \|q\mathcal A_{1,\parallel}\|/\partial t$ vanishes using the polarization equation. Then we have
\begin{align}\label{eq:parallel_momentum_direction}
&\text{for }\eta \in \{\varphi, \vartheta\}&  \nonumber\\
    &\frac{\partial}{\partial t} \sumsp  \RA{mnu_{\parallel} b_\eta}  
    \nonumber\\
    &+ \sumsp\frac{\partial}{\partial v} \RA{mnu_\parallel b_\eta u_E^\tv +  (p_\parallel + mnu_\parallel^2) b_\eta b_{1,\perp}^\tv } 
    \nonumber\\
   &= \sumsp - \RA{ p_\perp \frac{\partial \ln B}{\partial \eta} +(p_\parallel + mnu_\parallel^2 )b^i \frac{\partial b_i}{\partial \eta}} \nonumber\\
   &+\RA{(\vec j_f \times \vec B)_\eta} +\sumsp m\RA{ S_{nu_\parallel} b_\eta}
\end{align}
while the average parallel momentum reads
\begin{align}
       & \sumsp \left\{\frac{\partial}{\partial t}\RA{mnu_{\parallel} }
    + \frac{\partial}{\partial v} \RA{mnu_\parallel u_E^\tv + (p_\parallel + mnu_\parallel^2) b_{1,\perp}^\tv }  \right\}
    \nonumber\\
   &= \sumsp\left\{\RA{-p_\perp\npar \ln B} + m \RA{S_{nu_\parallel}}\right\}
   \label{eq:parallel_momentum}
\end{align}

The two components of Eq.~\eqref{eq:parallel_momentum_direction}
complement the previously derived \ExB velocity components in Eqs.~\eqref{eq:reynolds_flow_vartheta} and \eqref{eq:reynolds_flow}.

In total, Eqs~\eqref{eq:reynolds_flow}, \eqref{eq:reynolds_flow_vartheta}, \eqref{eq:parallel_momentum_direction}  and \eqref{eq:parallel_momentum} form the basis of our discussion for the remainder of the manuscript.

\section{Favre averaged momentum equations} \label{sec:momentum_fa}
In order to discuss the effect of turbulent fluctuations on flux-surface averaged
quantities a Reynolds decomposition is traditionally used to rewriteß nonlinearities in the averaged evolution equations. For any function $h(\vec x)$ we have
\begin{align}\label{eq:RA}
    h &\equiv \RA{h} + \RF{ h}
\end{align}
Unfortunately, as we point out in Reference~\cite{Held2018} the
Reynolds decomposition technique does not lead to well-behaved terms when the absolute density $n$ appears in the nonlinear terms in the sense that (i) absolute density fluctuations $\RF{n}$ appear instead of relative density fluctuations $\RF{n}/\RA{n}$, (ii) the radial advective part is not correctly recovered and (iii) effects from the density gradient $\partial_v \ln \RA{n}$ are not evident. We will thus follow~\cite{Held2018} and introduce the so-called Favre decomposition.

Consider a term of the form $\RA{nh}$.
If we
multiply and divide by $\RA{n}$, we can write
$\RA{nh} \equiv \RA{n}\FA{h}$.
Here, we introduce the so-called Favre average 
\begin{align}\label{eq:FA}
    \FA{h} := \frac{\RA{nh}}{\RA{n}} 
\end{align}
which can be understood as a density weighted Reynolds average. We note that this definition is species dependent through the dependence on the species density $n$.
The Favre average then allows the definition of the Favre decomposition
\begin{align}\label{eq:FF}
    h \equiv \FA{h} + \FF{ h}
\end{align}
The Favre average reduces to the Reynolds average for small fluctuation amplitudes or if the density is a flux-function $\FA{h}=\RA{h} + \RA{\RF{n}\RF{h}}/\RA{n} \approx\RA{h}$.  Reference~\cite{Held2018} also reported $\FA{u_\vartheta}\approx \RA{u_\vartheta}$ within a few percent since $\FA{\RF{ u_\vartheta}} \approx 0$ in gyro-fluid simulations.
We emphasize that the Favre average is
a technique to present an equation in a way that can be easily interpreted physically. While it changes the appearance of an equation it does not change its content.

\subsection{Favre averaged covariant \texorpdfstring{\ExB}{TEXT} velocity}
We first apply the Favre average technique to the continuity equation
$\partial_t n + \nc( n\vec u) = S_n$ to get
\highlight{
\begin{align}
%
  \label{eq:density_FA}
    \frac{\partial }{\partial t} \RA{n} 
    &+ \frac{\partial }{\partial v} \left(\RA{n}\mathcal U^v \right) =\RA{S_n}
\end{align}
where we define the average radial velocity
\begin{align}
    \mathcal U^v &:= \FA{u_E^\tv+u_{\parallel} b_{1,\perp}^\tv}  \label{eq:def_V}
\end{align}
}
and we use $u^v = u_E^v + u_\parallel b_{1,\perp}^v +\mathcal O(\delta^3)$. 
If we now replace all terms of the form $\RA{nh}$
with $\RA{n}\FA{h}$ in Eq.~\eqref{eq:reynolds_flow}, then insert the continuity Eq.~\eqref{eq:density_FA} and use $\FA{gh} = \FA{g}\FA{h}+\FA{\FF g\FF h}$ (Eq.~\eqref{eq:FA_identities}) and $\FA{u_D^\tv} = \mathcal O (\delta^3)$, we get 
\begin{align}\label{eq:reynolds_flow_fa}
    \sumsp &\left\{m\RA{n}\left(\frac{\partial}{\partial t}  + \mathcal U^v\frac{\partial}{\partial v} \right)\FA{u_{E,\varphi}} \right\}
        \nonumber\\%
    =& -\frac{\partial}{\partial v }\TotalStress{\perp,\varphi}{\tv}-  \RA{\left(\vec j_f\times \vec B \right)_\varphi}+ \sumsp m\mathcal S_{u_{E,\varphi}}
\end{align}
and similarly in Eq.~\eqref{eq:reynolds_flow_vartheta} we get
\begin{align}
     \sumsp &\left\{m\RA{n}\left(\frac{\partial}{\partial t}
     +  \mathcal U^v\frac{\partial}{\partial v}\right) \FA{u_{E,\vartheta}} \right\} 
        \nonumber\\%
    =&  -\frac{\partial}{\partial v }\TotalStress{\perp,\vartheta}{\tv}-  \RA{\left(\vec j_f\times \vec B \right)_\vartheta}+ \sumsp m\mathcal S_{u_{E,\vartheta}} \nonumber\\
    &- \left( \sumsp  m\RA{n}\FA{u_{E,\vartheta}}\mathcal U^v +\Theta_\vartheta^\tv \right)\frac{\partial}{\partial v} \ln \iota
 \label{eq:reynolds_flow_vartheta_fa}
\end{align}
where we identify with $\RA{ B_{1,\perp}^v} =\mathcal O(\delta^3)$ and $\RA{\Mem^{,v}}=\mathcal O(\delta^3)$ in the \DK ordering
\begin{align}
\text{for }& \eta \in \{\varphi,\vartheta\} \nonumber\\
\TotalStress{\perp,\eta}{\tv} :=& \sumsp m\RA{n}\Tstress{\perp,\vartheta}{\tv} +\Mstress{\vartheta}{\tv}\\
    \Tstress{\perp,\eta}{\tv}:=& \underbrace{\FA{\FF{ u_{E,\eta}}\FF{ u_{E}^\tv}}}_{\FExBstress{\eta}{\tv}}
    + \underbrace{\FA{\FF{ u_{E,\eta}} \FF{ u_{D}^\tv} }}_{\Fdiastress{\eta}{\tv}}
    \underbrace{-\FA{u_{E,\eta}}\FA{u_{\parallel} b_{1,\perp}^\tv}}_{\Xstress{\eta}{\tv}} 
    \label{eq:exb_stresses}\\
    \Mstress{\eta}{\tv} :=& \underbrace{-\frac{1}{\mu_0}\RA{\RF{ B_{1,\perp,\eta}}  \RF{  B_{1,\perp}^\tv}}}_{\MBstress{\eta}{\tv}} + \underbrace{\RA{ \RF{ B_{1,\perp,\eta}}\;\RF{ \Mem^{,v}}}}_{\Wstress{\eta}{\tv}} \label{eq:maxwell_stress}\\
    \mathcal S_{u_{E,\eta}} :=& \RA{ \RF{S_n}\RF{u_{E,\eta}}}+\RA{S_n}\left(\RA{u_{E,\eta}}-\FA{u_{E,\eta}}\right)
    \label{eq:source_uEeta}
\end{align}
Note that the density $\RA{n}$ is species dependent and therefore we
cannot divide Eqs.~\eqref{eq:reynolds_flow_fa} and \eqref{eq:reynolds_flow_vartheta_fa} by $\RA{n}$. What is usually possible is to neglect the electron mass, which reduces the sum Eqs.~\eqref{eq:reynolds_flow_fa} and \eqref{eq:reynolds_flow_vartheta_fa} to a sum
over all ion species.

Equations~\eqref{eq:reynolds_flow_fa} and \eqref{eq:reynolds_flow_vartheta_fa} describe the
evolution of the Favre averaged covariant components of the \ExB velocity in general, not
necessarily axisymmetric magnetic field geometry up to third order in the \DK ordering. 
On the left hand side we find a radial advection term proportional to $\mathcal U^v$~\cite{Held2018}.
The first term on the right hand side is the total perpendicular stress $\TotalStress{\perp,\eta}{\tv}$, which consists of the perpendicular Favre stress $\Tstress{\perp,\eta}{\tv}$ and the Maxwell stress $ \Mstress{\eta}{\tv}$. We note here that we define the Favre stress as a kinematic stress ( "stress divided by mass density") with units m$^2$/s$^2$ as opposed to the Maxwell stress which has units of stress N/m$^2$.

The kinematic Favre stress $\Tstress{\perp,\eta}{\tv}$ contains the \ExB Favre stress $\FExBstress{\eta}{\tv}$.
As Reference~\cite{Held2018} points out the \ExB Favre stress $ \FExBstress{\eta}{\tv}$ can be written as
\begin{align}\label{eq:favre_as_reynolds}
    \FExBstress{\eta}{\tv} = \Rstress{\eta}{\tv} - \FA{\RF{u_{E,\eta}}} \FA{\RF{u_E^\tv}} + \RA{\RF{n}\RF{u_{E,\eta}}\RF{u_E^\tv}}/\RA{n}
\end{align}
where the \ExB  Reynolds stress is $\Rstress{\eta}{\tv} := \RA{\RF{u_{E,\eta}}\RF{ u_{E}^\tv}} $~\cite{Diamond1991} and the often neglected~\cite{Diamond2013} triple term appears on the right-hand side of Eq.~\eqref{eq:favre_as_reynolds}. An advantage of the Favre decomposition is that the density fluctuations are 
automatically contained as relative fluctuation levels as is evident in the triple term in Eq.~\eqref{eq:favre_as_reynolds}. 
An analogous identity to Eq.~\eqref{eq:favre_as_reynolds}
holds for the diamagnetic Favre stress $\Fdiastress{\eta}{\tv}$
\begin{align}\label{eq:favredia_as_reynolds}
    \Fdiastress{\eta}{\tv} = \Dstress{\eta}{\tv} - \FA{\RF{u_{E,\eta}}} \FA{\RF{u_D^\tv}} + \RA{\RF{n}\RF{u_{E,\eta}}\RF{u_D^\tv}}/\RA{n},
\end{align}
which
encompasses the diamagnetic Reynolds stress $\Dstress{\eta}{\tv}:= \RA{\RF{u_{E,\eta}}\RF{ u_{D}^\tv}} $~\cite{Madsen2017}. Note that the diamagnetic Favre stress is asymmetric in contrast to \ExB Favre stress. In this form the diamagnetic stress consist of the radial component of the diamagnetic velocity together with the $\eta$ component of the \ExB velocity.
This is a consequence of using the pressure equation to
evaluate the time-derivative of the diamagnetic velocity~\cite{Madsen2017}, which we have done using Eq.~\eqref{eq:onehalf}. Otherwise the  transpose of the diamagnetic stress consisting of the radial \ExB component and the $\eta$ component of the diamagnetic velocity appears~\cite{Smolyakov2000}. We elaborate further on different interpretations of the \ExB angular momentum density in Section~\ref{sec:emfields}. 
In addition to $\FExBstress{\eta}{\tv}$ and $\Fdiastress{\eta}{\tv}$ we find the stress term $\Xstress{\eta}{\tv}$ that appears for fluctuating magnetic field
$\RF{b_{1,\perp}^\tv}$. This term is in fact a remainder of the actual magnetic flutter Favre stress term $m\FA{\FF{u_{E,\eta}}\FF{ u_\parallel b_{1,\perp}^v}}$ that would appear, had
we not neglected the $A_{1,\parallel}$ nonlinearities in the Hamiltonian~\eqref{eq:hamiltonian} (through the variaton in Eq.~\eqref{eq:derivative_as_variation}).
We expect $\Xstress{\eta}{\tv}$ to vanish for small relative density fluctuations and to only play a role for \(\mathcal{O}(\RF{n}/\RA{n})\sim 1\) fluctuation amplitudes, due to the similar dependence as  the second term in the Favre stresses~\cite{Held2018}.

The Maxwell stress $\Mstress{\eta}{\tv}$ consists of the symmetric vacuum field contribution $\MBstress{\eta}{\tv}$ and the asymmetric magnetization stress term $\Wstress{\eta}{\tv}$. The role of the vacuum Maxwell stress $\MBstress{\eta}{\tv}$ on the generation of sheared \ExB flows was highlighted previously in for example~\cite{Craddock1991,Scott2005}.
The novel asymmetric magnetization stress $\Wstress{\eta}{\tv}$ appears in its present form analogously to the diamagnetic stress as a consequence of using the pressure equation~\eqref{eq:onehalf}. In Section~\ref{sec:emfields} we will encounter its transpose in the full electromagnetic field stress tensor. It notably contains a contribution
from the parallel heat flux $q_\parallel+ p_\perp u_\parallel$ and physically originates in the magnetization term in parallel Amp\`ere's law Eq.~\eqref{eq:induction_fluid}.

As was highlighted in~\cite{Held2018} the density gradient $\partial_v\ln\RA{n}$ contributes to the evolution of \ExB shear flow. Consider
\begin{align}\label{eq:derivativeTotalStress}
\frac{\partial}{\partial v}\TotalStress{\perp,\eta}{\tv} = \sumsp\left\{m\RA{n}\left(\frac{\partial}{\partial v}  \Tstress{\perp,\eta}{\tv} +  \Tstress{\perp\eta}{\tv} \frac{\partial}{\partial v}\ln\RA{n}\right)\right\}+ \frac{\partial}{\partial v}\Mstress{\eta}{\tv} 
\end{align}
 We emphasize that both \ExB and diamagnetic Favre stresses appear in the density gradient drive term $m\RA{n}\Tstress{\perp,\eta}{\tv} \partial_v\ln \RA{n}$ and that this term is non-zero even if $\partial_v \Tstress{\perp\eta}{\tv}$ vanishes.
This is particularly interesting
for the steep density gradient that develops during the transition to H-mode.

Contrary to the toroidal angular momentum density, the poloidal \ExB angular momentum density in Eq.~\eqref{eq:reynolds_flow_vartheta_fa} is influenced by a gradient in the rotational transform profile or magnetic shear $\sigma = \partial_v \ln \iota$.
The magnetic shear is known to influence the \ExB shear flow evolution~\cite{Burrell1997,Kendl2003}. In particular the shear dampens drift-wave turbulence and leads to narrow zonal flows~\cite{Kendl2003}. Furthermore, it dampens the Kelvin-Helmholtz instability, which would otherwise be driven by the \ExB velocity shear~\cite{Burrell1997}. In Eq.~\eqref{eq:reynolds_flow_vartheta_fa}, we explicitly identify two magnetic shear contributions.
The first shear term $m\RA{n}\mathcal U^v \FA{u_{E,\vartheta}} \sigma$ corresponds to roughly exponential growth or damping of poloidal flows, assuming that the average radial velocity is constant (which is a good estimate since it is approximately the \ExB radial particle transport). 
The second shear term appears analogous to the density gradient $\partial_v \ln n$ term and contributes even if \(\RA{n}\Tstress{\perp,\vartheta}{\tv}\) and $\Mstress{\vartheta}{\tv}$ are "radially" homogeneous (no volume derivative).

On the right hand side of Eq.~\eqref{eq:reynolds_flow_fa} and \eqref{eq:reynolds_flow_vartheta_fa} we further find the components of
the Lorentz force originating from the radial curvature drift current $\vec j_f$ defined in Eq.~\eqref{eq:free_current}. The grad-B induced current part of this term is the Stringer-Winsor spin-up term~\cite{Hassam1993,Hassam1994,Hallatschek2000}. In order to see this recall that $(\vec j_f \times \vec B)_\varphi = \vec j_f \cn\psi_p\sim p_\perp\mathcal  K(\psi_p)$ that is the radial component of the free current (see \ref{tab:operators} for the definition of the curvature operator $\mathcal K$).
The same term was found in $\delta F$ drift-fluid models~\cite{Scott2003, Scott2005, Naulin2005} and was there called the geodesic transfer term.
In any case the term is known to
excite geodesic acoustic modes and to both dampen or drive zonal flows depending on the parameter regime~\cite{Hallatschek2000,Scott2003}. We further discuss this term in relation to the ion orbit loss mechanism in Section~\ref{sec:ion-orbit-loss}.

Finally, on the right side of Eq.~\eqref{eq:reynolds_flow_fa} and \eqref{eq:reynolds_flow_vartheta_fa} we find source related terms contained
in $\mathcal S_{u_{E,\eta}}$ defined in Eq.~\eqref{eq:source_uEeta}.
The term $\RA{\RF{S_n}\RF{u_{E,\vartheta}}}$ in Eq.~\eqref{eq:reynolds_flow_vartheta_fa} describes the poloidal spin-up mechanism for poloidally asymmetric particle sources described in~\cite{Hassam1994}. In Eq.~\eqref{eq:reynolds_flow_fa}
we find an equivalent term also for the toroidal \ExB velocity. A poloidally (or toroidally)
asymmetric particle source can generate or dampen toroidal \ExB velocity. This should be contrasted with Reference~\cite{Helander2003}, which finds angular momentum generation susceptible to the poloidal location of neutrals through viscosity and heat flux effects. In this contribution collisional effects are treated only indirectly subsuming the collision operator into the kinetic "source" term $S$ in the Vlasov equation~\eqref{eq:vlasov}.
The second source term is proportional to the difference between Reynolds and Favre averaged \ExB velocity $\RA{u_{E,\eta}} - \FA{u_{E,\eta}}$. For small density
fluctuations we thus expect this term to vanish and only contribute for large
fluctuation amplitudes.

\subsection{Favre averaged parallel velocity}
For the parallel angular momentum components~\eqref{eq:parallel_momentum_direction} we have
\begin{align}\label{eq:parallel_momentum_direction_fa}
&\text{for }\eta \in \{\varphi, \vartheta\}&  \nonumber\\
    \sumsp&\left\{m\RA{n}\left(\frac{\partial}{\partial t} \FA{u_{\parallel}b_\eta }
    +\mathcal U^v\frac{\partial}{\partial v}\FA{u_{\parallel}b_\eta}\right)\right\}
    \nonumber\\
    = &-\frac{\partial}{\partial v}\TotalStress{\parallel,\eta}{\tv}+ 
    \sumsp \RA{ -p_\perp \frac{\partial \ln B}{\partial \eta} -(p_\parallel + mnu_\parallel^2 )b^i \frac{\partial b_i}{\partial \eta}} \nonumber\\
   &+\RA{(\vec j_f \times \vec B)_\eta} 
   +\sumsp m\mathcal S_{u_\parallel b_\eta}
\end{align}
where $\mathcal U^v$ is given in Eq.~\eqref{eq:def_V} and we identify
\begin{align}
\TotalStress{\parallel,\eta}{\tv} :=& \sumsp \left\{m\RA{n} \Tstress{\parallel,\eta}{\tv} + \Kstress{\parallel,\eta}{\tv}\right\} \\
\Kstress{\parallel,\eta}{\tv}:=&\RA{\RF{p_{\parallel}b_\eta}\; \RF{ b_{1,\perp}^\tv} } \label{eq:kinetic_stress}\\
\Tstress{\parallel,\eta}{\tv} :=&
\underbrace{\FA{\FF{u_{\parallel}b_\eta}\; \FF{u_{E}^\tv}} }_{\FExBstress{\parallel,\eta}{\tv}}
+ \underbrace{ \FA{\FF{u_{\parallel}b_\eta} \; \FF{u_{\parallel}b_{1,\perp}^\tv} } }_{\Hstress{\parallel,\eta}{\tv}} \label{eq:parallelfavrestress}
 \\
\mathcal S_{u_\parallel b_\eta}:=&\RA{S_{nu_\parallel} b_\eta}-\RA{S_n}\FA{u_\parallel b_\eta}
\end{align}
With the Favre average we re-write Eq.~\eqref{eq:parallel_momentum} into
\begin{align}\label{eq:parallel_momentum_fa}
    \sumsp& \left\{m\RA{n}\left(\frac{\partial}{\partial t} \FA{u_{\parallel} }
    +\mathcal U^v\frac{\partial}{\partial v}\FA{u_{\parallel}}\right)\right\}
    \nonumber\\
    =&-\frac{\partial}{\partial v}\TotalStress{\parallel}{\tv}+\sumsp\left\{m\mathcal S_{u_\parallel}-
    \RA{p_\perp \npar\ln B}\right\}
\end{align}
where $\mathcal U^v$ is given in Eq.~\eqref{eq:def_V} and we identify
\begin{align}
\TotalStress{\parallel}{\tv} :=& \sumsp \left\{m\RA{n} \Tstress{\parallel}{\tv} + \Kstress{\parallel}{\tv}\right\} \\
\Kstress{\parallel}{\tv}:=& \RA{\RF{p_{\parallel}}\; \RF{b_{1,\perp}^\tv}}\\
\Tstress{\parallel}{\tv} :=&
\underbrace{ \FA{\FF{u_{\parallel}}\; \FF{u_{E}^\tv}} }_{\FExBstress{\parallel}{\tv}}
+ \underbrace{ \FA{\FF{u_{\parallel}} \; \FF{u_{\parallel}b_{1,\perp}^\tv} } }_{\Hstress{\parallel}{\tv}} \\
\mathcal S_{u_\parallel} :=& \RA{S_{nu_\parallel} }-\RA{S_n}\FA{u_\parallel}
\end{align}
Analogous to Eq.~\eqref{eq:reynolds_flow_fa} in Eqs~\eqref{eq:parallel_momentum_fa} and \eqref{eq:parallel_momentum_direction_fa}
 we find a radial
advection term of momentum by $\mathcal U^v$ followed by various stress terms contained in $\TotalStress{\parallel}{v}$ respectively $\TotalStress{\parallel,\eta}{\tv}$. Again, we define the Favre stress as a kinematic stress and analogous relations to Eq.~\eqref{eq:favre_as_reynolds} hold for the parallel Favre
stress components. The parallel \ExB Favre stress $\FExBstress{\parallel}{\tv}$ respectively $\FExBstress{\parallel,\eta}{\tv}$
now depends on fluctuations in the parallel velocity instead of \ExB velocity. The Reynolds stress analogue of $\FExBstress{\parallel}{\tv}$ is well-known in the literature on intrinsic toroidal rotation (see e.g.~\cite{Diamond2013}), however we point out here that $\FExBstress{\parallel,\eta}{\tv}$ is the actual component that drives angular momentum $\FA{u_\parallel b_\eta}$ instead of just parallel momentum $\FA{u_\parallel}$.
The parallel magnetic flutter Favre stress term $\Hstress{\parallel}{\tv}$ respectively $\Hstress{\parallel,\eta}{\tv}$ is a transfer term appearing for magnetic fluctuations $b_{1,\perp}$. 
The kinetic stress term $\Kstress{\parallel}{\tv}$ respectively $\Kstress{\parallel,\eta}{\tv}$ is related to the kinetic dynamo mechanism as for example discussed for the reversed field pinch in Reference~\cite{Prager1999,Ding2013}.  On the right hand side we find the mirror force term $-\RA{p_\perp \npar\ln B}$ respectively $-\RA{b_\eta p_\perp \npar\ln B}$. In the equation for the parallel angular momentum Eq.~\eqref{eq:parallel_momentum_direction_fa} we find an additional geometrical correction to the mirror force. Finally, the momentum source term $\mathcal S_{u_\parallel}$ respectively
$\mathcal S_{ u_\parallel b_\eta}$ represents angular momentum generation by external sources. Note that with the definition of a velocity source $S_{u_\parallel}$ via $S_{nu_\parallel}:= n S_{u_\parallel}+ u_\parallel S_n$ we can write 
\begin{align} \label{eq:source_parallel_fluc}
    \mathcal S_{u_\parallel} = \RA{n}\FA{S_{u_\parallel}} + \RA{\RF{u_\parallel}\RF{S_n}} + \RA{S_n}\left(\RA{u_\parallel} - \FA{u_\parallel}\right)
\end{align}
and analogous for $\mathcal S_{u_\parallel b_\eta}$. Eq.~\eqref{eq:source_parallel_fluc} now consists of the Favre averaged velocity source plus a contribution from a poloidally asymmetric source term analogous to Eq.~\eqref{eq:source_uEeta}.

We comment here on the appearance of the Lorentz force in the equation for the parallel angular momentum Eq.~\eqref{eq:parallel_momentum_direction_fa}.
The Lorentz force acts perpendicularly to the magnetic field line 
and should not  contribute to the parallel momentum at all.
Indeed, we can further simplify the right hand side of Eq.~\eqref{eq:parallel_momentum_direction_fa} to
\begin{align}\label{eq:parallel_lorentz}
    \sumsp& \RA{ -p_\perp \frac{\partial \ln B}{\partial \eta} -(p_\parallel + mnu_\parallel^2 )b^i \frac{\partial b_i}{\partial \eta}}
   +\RA{(\vec j_f \times \vec B)_\eta} 
   \nonumber\\
   &= -\sumsp \RA{ p_\perp b_\eta\nabla_\parallel\ln B +(p_\parallel+mnu_\parallel^2) \tilde\kappa_\eta}
\end{align}
where we define $\tilde\kappa_\eta:=\eeta \cdot \vec \kappa - b^i\partial_\eta b_i$ with the curvature $\vec \kappa :=\bhat \cn \bhat$. Now, only the component of the mirror force $-p_\perp b_\eta\nabla_\parallel \ln B$ and a geometric correction term appear.
To see the mirror force recall the sign of the magnetic moment vector $\vec\mu = - \mu \bhat$ and the guiding centre parallel magnetization density~\cite{BrizardHahm,Brizard2013} $\vMgy_\parallel = \|\vec \mu\| = -\|\mu\| \bhat = -P_\perp \bhat /B$. 
The force acting on
magnetic dipoles is~\cite{Jackson} $\vec f_d =\vn( \vec\mu\cdot \vec B) = - \mu \vn B$. Taking the velocity space moment
we get $\vec F_d = \|\vec f_d\| = M_\parallel\vn B = -P_\perp \vn \ln B = -P_\perp \bhat\npar \ln B -P_\perp \np \ln B$. 
The parallel part reads $F_{d,\parallel} = -P_\perp \npar\ln B $, which is what appears in Eq.~\eqref{eq:parallel_momentum}, while $F_{d,\eta} = -P_\perp \partial_\eta \ln B$ appears in \eqref{eq:parallel_momentum_direction}. The perpendicular part gives rise to the $\vn B$ drift. Last, notice that $\npar \ln B = -\nc\bhat$ such that
\begin{align}
    -\RA{p_\perp \npar\ln B} &= \RA{\npar \RF{p_\perp}} \\
    -\RA{b_\eta p_\perp \npar\ln B} &= \RA{\npar (\RF{b_\eta p_\perp})}
\end{align}
Pressure fluctuations are required to affect the the angular momentum generation via the mirror force.

\subsection{Total angular momentum density}
The velocity equations~\eqref{eq:reynolds_flow_fa}/\eqref{eq:reynolds_flow_vartheta_fa} and \eqref{eq:parallel_momentum_direction_fa} can be easily cast back into conservative form using
 the continuity equation~\eqref{eq:density_FA} and $\RA{n}\FA{h} = \RA{nh}$ for any $h$.
Summing up the results, we finally find the evolution of the total average poloidal and toroidal angular momentum density
\begin{align}\label{eq:angular_momentum_full}
&\text{for }\eta \in \{\varphi, \vartheta\}&  \nonumber\\
\sumsp& \left\{ \frac{\partial}{\partial t} m\RA{n(u_{\parallel}b_\eta + u_{E,\eta})} \right.
\nonumber\\
&+\left.\frac{\partial}{\partial v}  m\RA{n(u_{\parallel}b_\eta + u_{E,\eta})} \mathcal U^v\right\}
+  \frac{\partial}{\partial v}\left(\TotalStress{\perp,\eta}{\tv}+\TotalStress{\parallel\eta}{\tv} \right)
\nonumber\\
=& -\delta_{\eta\vartheta} \left( \sumsp  m\RA{nu_{E,\vartheta}}\mathcal U^v +\TotalStress{\perp,\vartheta}{\tv}+\TotalStress{\parallel\vartheta}{\tv} \right)\frac{\partial}{\partial v}\ln \iota  
    \nonumber\\
    - \sumsp& \RA{ p_\perp \frac{\partial \ln B}{\partial \eta} +(p_\parallel + mnu_\parallel^2 )b^i \frac{\partial b_i}{\partial \eta}
    -m\left( S_{nu_\parallel} b_\eta + S_n u_{E,\eta}\right)}
\end{align}
where  $\delta$ is the Kronecker delta. The magnetic shear term only contributes to the poloidal angular momentum. The convective
term proportional to $\mathcal U^v$  vanishes under volume integration up to a surface contribution as does the
total stress term $\TotalStress{\perp,\eta}{\tv}+\TotalStress{\parallel\eta}{\tv}$.
In Eq.~\eqref{eq:angular_momentum_full} we further find that the momentum transfer to the background magnetic field is mediated by the mirror force and the generalized curvature force term on the right hand side.  Clearly, the Lorentz force term cancels in the total angular momentum density evolution. Finally, we recover the external source terms on the right hand side. 

We see that the total angular momentum in Eq.~\eqref{eq:angular_momentum_full} is given by the covariant components of the \ExB and parallel velocities. Comparing this 
to the total advection velocity $\vec u := \|\dot{\vec X}\| = \vec u_E + \vec u_\kappa + \vec u_{\vn B} + u_\parallel \bhat  + u_\parallel \vec b_{1,\perp}$ that appears in the continuity equation $\partial_t n + \nc (n\vec u) = S_n$ we see that the curvature, grad-B and
magnetic flutter velocities do not appear in the angular momentum~\eqref{eq:angular_momentum_full}
even though we at least expected the magnetic flutter term as an order $\mathcal O( \delta)$ term.
At this point recall Eq.~\eqref{eq:simplified_psipdot}, which
identifies the radial polarization current with the macroscopic expression for the angular momentum density (except $u_\parallel b_\eta$). The polarization density $\vPgy$ is directly connected to the definition of the Hamiltonian~\eqref{eq:hamiltonian} through the variational principle. Since we neglected the second order guiding center corrections we accordingly miss the guiding center polarization density~\cite{Brizard2011,Brizard2013} and
thus the corresponding curvature terms in our angular momentum density. On the other hand we also neglected the nonlinear terms in $A_{1,\parallel}$ in the Hamiltonian, which accounts for the missing magnetic flutter velocity $mv_\parallel b_{1,\perp}$ in the polarization~\cite{BrizardHahm} and thus angular momentum density~\eqref{eq:angular_momentum_full}.  

\section{The rotational energy}\label{sec:rotational}
\subsection{Angular momentum and angular velocity}
In Section~\ref{sec:momentum_fa} we have derived equations for the 
covariant components of the \ExB and parallel velocity, which add up to the total angular momentum density in Eq.~\eqref{eq:angular_momentum_full}. We now focus on the angular momentum as a vector quantity. We define
\begin{align}
    \vec u_L := \vec u_E  + u_\parallel\bhat = u_{L,\varphi } \vn\varphi + u_{L,\vartheta }  \vn\vartheta + u_{L,v} \vn v
\end{align}

We are now interested only
in the part of the flow that stays within a given flux-surface, because this flow can be constructed from the covariant $\varphi$ and $\vartheta$ components of $\vec u_L$ that we have available. 
To see this, we formulate the projection tensor
onto the flux surfaces 
\begin{align}
h_S:= \mathbf{1}-\rhohat\rhohat
\end{align}
with the contravariant radial unit vector $\rhohat := \vn v/|\vn v|$. With this we can split the flow velocity
according to $\vec u_L = \vec {u}_L|_{\psi_p}+ u_L^{\rhohat} \rhohat$ where
we define the surface or rotational velocity
\begin{align}\label{eq:surface_flow}
    \vec L \equiv \vec u_L|_{\psi_p} := h_S \cdot\vec u_L =& u_{L,\vartheta}\vn_S \vartheta + u_{L,\varphi}\vn_S\varphi \nonumber\\
    =& u_{L}^\vartheta \etheta + u_L^\varphi \ephi 
\end{align}
where we follow~\cite{Grimm1983} and introduce the surface operator $\vn_S := h_S\cdot \vn$. We thus have $L_i = u_{E,i} + u_\parallel b_i$ for $i\in\{\varphi,\vartheta, \rho\}$. As expected we do not need the radial component of $\vec u_L$ to construct the surface flow in Eq.~\eqref{eq:surface_flow}.

It is now important to see
that $\vn_S \varphi$ and $\vn_S\vartheta$
form the contravariant basis of the flux surface as a stand-alone manifold and analogous $\ephi$ and $\etheta$ are its covariant basis vectors
 In fact, explicitly writing $h_S$ into components
we realize that all components $h_{S,\rho k}$ vanish for $k\in \{\varphi,\vartheta,\rho\}$.
We thus define $\form$ as the two-dimensional 
tensor consisting of the non-zero components of $h_S$, that is
\begin{align}\label{eq:fundamental}
    \form := \begin{pmatrix}
     g_{\vartheta\vartheta} & g_{\vartheta\varphi} \\
     g_{\varphi\vartheta} & g_{\varphi\varphi}
    \end{pmatrix}
\end{align}
The interested reader will recognize $\form$ as the the first fundamental form of flux surfaces parameterized with $\vartheta$ and $\varphi$.
The first fundamental form $\form$ can be interpreted as the two-dimensional metric tensor of the flux-surface thought as a standalone structure and is thus an intrinsic structure of the magnetic flux surfaces (and in particular has a well-defined expression in every coordinate system).
Unfortunately, the flux-surface average is not an intrinsic surface operation since it requires the knowledge of the volume form $\sqrt{g}$ to compute.
Also, note that the components of $\form$ and its inverse are given by $\form_{\varphi\varphi} = \ephi\cdot\ephi$, $\form_{\vartheta\varphi} = \etheta\cdot\ephi$, $\form_{\vartheta\vartheta} = \etheta\cdot\etheta$ and $\form^{\varphi\varphi} = \vn_S\varphi\cdot\vn_S\varphi$, $\form^{\vartheta\varphi} = \vn_S\vartheta\cdot\vn_S\varphi$, $\form^{\vartheta\vartheta} = \vn_S\vartheta\cdot\vn_S\vartheta$ respectively.

Now, the fundamental form $\form$ has another interpretation, namely as the inertia tensor of rotations in $\vartheta$ and $\varphi$. To see this recall that the \textit{contravariant} components of the
surface velocity $\urho$, $L^\varphi$ and $L^\vartheta$ are actually
the \textit{angular velocities} with units s$^{-1}$, because the particle trajectory is given by $\dot\varphi = L^\varphi$ and $\dot \vartheta = L^\vartheta$. In contrast, the \textit{covariant} components 
$L_i=\form_{ij}L^j$ for $i,j\in\{\vartheta,\varphi\}$ form the \textit{angular momentum} as it results in Eq.~\eqref{eq:angular_momentum_full} that is $mL_i$ has units kgm$^2$s$^{-1}$.
This leaves $m\form$ as the (kinematic) inertia tensor that connects the angular velocity and angular momentum of a fluid element rotating on a flux-surface.
\subsection{Mean and fluctuating angular momentum}
Consider now the mean surface velocity field generated by Favre averaged covariant $\varphi$ and $\vartheta$ velocity components
\begin{align}\label{eq:mean_flow}
    \vec L_m := \FA{L_{\vartheta}}\vn_S\vartheta + \FA{L_\varphi} \vn_S\varphi 
\end{align}
The time evolution of $m\RA{n}\vec L_m$ is directly given by Eq.~\eqref{eq:angular_momentum_full}.
First, we emphasize that the corresponding \textit{angular velocity} components of $\vec L_m$, $L^i = \form^{ij} \FA{L_i}$ are
\textit{not} flux functions since the inertia tensor does not commute with the flux-surface average and thus $\vec L_m \neq \FA{L^\vartheta}\vec e_\vartheta + \FA{L^\varphi}\vec e_\varphi$
or in other words, if angular momentum is a flux-function then angular velocity cannot be at the same time. In fact, we perform the splitting $L_i = \FA{L_i}+\FF{L_i}$ expecting that the relative fluctuations $\FF{L_i}$ are small and that $u_i$ is well-described by its Favre average $\FA{L_i}$.
A priori, these arguments of course also hold the other way, if angular velocity were a flux-function then angular momentum cannot be at the same time and we should split the angular velocities.

At this point recall the discussion in the introduction. 
When angular momentum is conserved, a particle moves faster closer to the axis (for example on the high field side in Fig.~\ref{fig:flux_grid}). We take this as an indication that angular velocities are not well-described by flux-surface averages, while angular momenta are.
Furthermore, in the equations in Section~\ref{sec:momentum_fa} (for example Eq.~\eqref{eq:reynolds_flow_fa}) we see
that the average \textit{angular momentum}
is fed by turbulent fluctuations through the stress tensor, which we interpret as an indication that fluctuations $\FF{L_i}$ and not $\FF{L^i}$ become small. 

\subsection{Total energy evolution}
Before we construct a zonal or mean flow rotational energy we first focus on the total energy evolution of our system.
We 
follow Reference~\cite{Madsen2015} and derive the pressure
equations (the thermal energy) for $p_\perp$ and $p_\parallel$ directly from the moment evolution equation~\eqref{eq:zeta_general}.
We point out that we need to keep terms one order higher in the energy conservation law than in the momentum conservation law, that is $\mathcal O(\delta^4)$ in our ordering. 
This is due to the fundamental property of the gyro-kinetic system~\cite{BrizardHahm} that a higher order Hamiltonian needs to be kept in the system to obtain polarization effects and an exact energy invariant. If we thus neglect all terms
of order $\mathcal O(\delta^5)$, use parallel Amp\`ere's law~\eqref{eq:induction_fluid} and apply the species summation we get
\begin{align}\label{eq:energy_pressure}
     \frac{\partial}{\partial t}& \RA{ \sumsp\left\{ p_\perp + \frac{1}{2}p_\parallel + \frac{1}{2}mnu_\parallel^2\right\} + \frac{(\np A_{1\parallel})^2}{2\mu_0}}
     \nonumber\\
     +&\frac{\partial}{\partial v} \RA{ j_{\mathcal E,p}^\tv}
   = \RA{\vec j_f \cdot\vec E_\perp + j_\parallel E_\parallel} + \sumsp \RA{S_{p_\perp} + \frac{1}{2}S_{ p_\parallel + mnu_\parallel^2}}
\end{align}
where $\vec E_\perp= -\np\phi$ and $E_\parallel = - (\npar \phi + \vec b_{1,\perp}\cn\phi)$ and $j_\parallel := \sumsp qnu_\parallel$.
We formally summarize all total divergences into the term $j_{\mathcal E,p}^\tv$.
 An interesting side-remark here is to view the energy conservation Eq.~\eqref{eq:energy_pressure} to lowest order, which leaves Bernoulli's identity $\RA{p_\perp + p_\parallel/2 + mu_\parallel^2/2} = const$ along fluid trajectories.
On the right side of Eq.~\eqref{eq:energy_pressure} appears the energy exchange term $ \RA{\vec j_f\cdot\vec E_\perp + j_\parallel E_\parallel}$ as well as the pressure source terms (heating).

On the other side using the definition of $\Psi$ in Eq.~\eqref{eq:generalized_potential_phi} and the polarization equation~\eqref{eq:polarization_fluid} we find
\begin{align}\label{eq:recover_ExB} 
    \sumsp \|q\Psi\| = \sumsp \nc \left( \frac{m\|\mu B\|}{2qB^2}\np \phi -\phi\np \frac{m\|\mu B\|}{2qB^2}\right) \nonumber\\
    - \vn \left(\phi \frac{ mN\np\phi}{B^2} \right) 
    +\frac{1}{2}mN \frac{(\np\phi)^2}{B^2}
\end{align}
which recovers the \ExB kinetic energy density in the last term on the right hand side.
Interestingly, a completely analogous relation holds for the term $\|q\Psi\|_S$ (by replacing $\|\mu B\|$ with $\| \mu B\|_S$ and $N$ with $S_N$ in Eq.~\eqref{eq:recover_ExB}) since
we require the sources to preserve quasineutrality in Eq.~\eqref{eq:quasineutral_sources}.
Applying Eq.~\eqref{eq:zeta_general} to $q\dot \Psi = q\partial_t \Psi + q\vec{\dot X}\cn \Psi $ and using~\eqref{eq:recover_ExB} and \eqref{eq:etagradH} for $\Psi$ under species summation
and neglecting again terms of order $\mathcal O(\delta^5)$ the result is given by
\begin{align}\label{eq:ExBenergy}
    \frac{\partial}{\partial t} \RA{\frac{1}{2}\rho_M u_E^2} + \frac{\partial}{\partial v} \RA{j_{\mathcal E, \psi}^\tv}
    = -\RA{\vec j_f \cdot \vec E_\perp + j_\parallel E_\parallel} \nonumber\\
    + \sumsp \frac{1}{2} m \RA{S_nu_E^2 }
\end{align}
where we identify the total mass density $\rho_M := \sumsp mn$ since the \ExB velocity is species independent and again summarize all divergences into the formal $j_{\mathcal E,\psi}^\tv$ term. The density source $S_n$ either generates or destroys kinetic \ExB energy depending on its sign. The term appears analogous to the momentum source in Eq.~\eqref{eq:reynolds_flow}. The sum of Eqs.~\eqref{eq:energy_pressure} and \eqref{eq:ExBenergy} recovers the conservation of the flux-surface averaged total energy of our model since the energy exchange term $ \RA{\vec j_f\cdot\vec E_\perp + j_\parallel E_\parallel}$ cancels.
 \subsection{Mean rotational energy evolution}

The direct approach to a rotational energy density is the kinetic energy of the surface flow velocity $\vec L$
\begin{align}\label{eq:rotational_energy}
     E_{\mathrm{rot}} := \sumsp \frac{1}{2} m\RA{n \vec L \cdot \form \vec L}=   \RA{\frac{1}{2}\rho_M\vec u_E^2|_{\psi_p}} +\sumsp \RA{\frac{1}{2}mn u_\parallel^2}
\end{align}
This energy is equivalent to subtracting the radial \ExB energy $\RA{\rho_M u_{E,v} u_E^v/2}$ from the total kinetic energy density $\sumsp  \RA{mn (\vec u_E^2 + u_\parallel^2)/2}$. It is now important to realize that contrary to the parallel kinetic energy the \ExB rotational energy density can be related to the (species summed) angular momentum evolution. This is because the \ExB drift velocity is equal for all species.
%
We can write
\begin{align}
       \frac{1}{2}\RA{\rho_M \vec u_E|_{\psi_p}^2}
    = E_{\mathrm{zonal}} +  E_{\mathrm{fluc}}
\end{align}
where we define
\begin{align}
    E_{\mathrm{zonal}} :=&
    \frac{1}{2}\RA{\rho_M}  \FMA{\form^{ij} } \FMA{u_{E,i}}\FMA{u_{E,j}} \\
     E_{\mathrm{fluc}} :=& \RA{\rho_M}
    \FMA{\form^{ij}\FMA{u_{E,i}}\FF{ u_{E,j}}} \nonumber\\
     &+
     \frac{1}{2}\RA{\rho_M}\FMA{\form^{ij}} \FMA{\FF{ u_{E,i}} \FF{ u_{E,j}}}
\end{align}
and here introduce the total mass density in the Favre averages
\begin{align}\label{eq:FMA}
    \FMA{h} := \RA{\rho_M}^{-1}\sumsp m \RA{nh}
\end{align}
for any (possibly species dependent) function $h$. If $h$ is species independent Eq.~\eqref{eq:FMA} simplifies to $\FMA{h}=\RA{\rho_M h}/\RA{\rho_M}$.
With $ u_{E,\vartheta} = \iota^{-1} u_{E,\varphi} $ we can simplify further
\begin{align}
    E_\mathrm{zonal} &= \frac{1}{2}\RA{\rho_M}\FMA{ \iota^{-2}\form^{\vartheta\vartheta} + 2\iota^{-1}\form^{\vartheta\varphi} + \form^{\varphi\varphi}} \FMA{u_{E,\varphi}}^2
    \nonumber\\
    &\equiv \frac{1}{2} \RA{\rho_M} \FMA{u_{E,\varphi}}^2  \FMA{\form_0}
    \label{eq:zonal_energy}
\end{align}
Here, we introduce the inertia factor $\form_0 := (\iota^{-1},1)\form^{-1} (\iota^{-1},1)^\mathrm{T}$. For a purely toroidal magnetic field we have $\form_0 = R^{-2}$ as expected.
For symmetry flux coordinates we have $g_{\vartheta\vartheta} = R^2 (\vn\psi_p)^2/I^2\iota^2$, $g_{\varphi\vartheta} =0$ and $g_{\varphi\varphi}=R^2$ and thus $\form_0 = R^{-2}( 1 + I^2/|\vn\psi_p|^2)= B^2 / |\vn\psi_p|^2$. The inertia factor vanishes for a slab magnetic field. In this case our zonal flow energy agrees with~\cite{Held2018} and in the case of small density fluctuations also with its $\delta F$ analogue~\cite{Scott2005,Naulin2005}.
Since $\form_0$ is time-independent we can use the evolution equations for the density Eq.~\eqref{eq:density_FA} and angular momentum~\eqref{eq:reynolds_flow_fa} to get
\begin{align}\label{eq:perp_kinetic}
      \frac{\partial}{\partial t}&E_{\mathrm{zonal}} +\frac{\partial}{\partial v } \left(E_{\mathrm{zonal}}\FMA{u^v} \right)
  \nonumber\\
  =&-\FMA{\form_0}\FMA{u_{E,\varphi}}\left(\frac{\partial }{\partial v}  \TotalStress{\perp,\varphi}{\tv} + \RA{(\vec j_f\times\vec B)_\varphi}\right)
  \nonumber\\
    &-\frac{1}{2}\FMA{u_{E,\varphi}}^2\frac{\partial}{\partial v}\left(\RA{\rho_M}\FMA{\FF{\form_0}\FF{u^v}}\right)
     + \mathcal S_{\mathrm{zonal}}
\end{align}
where we neglected the term $\RA{n\vec u\cn \form_0}$ in the continuity equation as small in our ordering
 and we have 
 \begin{align}
 \mathcal S_{\mathrm{zonal}} :=& \FMA{\form_0} \FMA{u_{E,\varphi}} \mathcal S_{u_{E,\varphi}} + \frac{1}{2}\FMA{u_{E,\varphi}}^2  \sumsp  \RA{mS_n\form_0}
 \label{eq:zonal_source}
 \end{align}


In Eq.~\eqref{eq:perp_kinetic} we find the term $\FMA{u^v}$ as the convective velocity for the zonal flow energy. 
 On the right side the derivative of the total perpendicular stress $\TotalStress{\perp,\varphi}{\tv}$ given by Eq.~\eqref{eq:derivativeTotalStress} appears. Thus, the Favre and Maxwell stress given by fluctuating velocities and the fluctuating magnetic field in Eqs.~\eqref{eq:exb_stresses} and \eqref{eq:maxwell_stress} respectively together with a gradient in the density $\partial_v \ln \RA{n}$ can appear as
sources for zonal flow energy. The \ExB Favre stress $\FExBstress{\varphi}{\tv}$ was already identified as  a source for zonal flow energy in a slab geometry in ~\cite{Held2018}.  The vacuum field Maxwell stress $\MBstress{\varphi}{\tv}$ and the \ExB Reynolds stress $\Rstress{\varphi}{\tv}$ (contained in our $\Tstress{\perp,\varphi}{\tv}$ according to Eq.~\eqref{eq:favre_as_reynolds}) appear in similar form in $\delta F$ models~\cite{Scott2005,Naulin2005}. Compared to these previous findings we find the additional appearance of the diamagnetic Favre stress $\Fdiastress{\varphi}{\tv}$ contained in $\Tstress{\perp,\eta}{\tv}$ and the magnetization stress $\Wstress{\varphi}{\tv}$ contained in $\Mstress{\varphi}{\tv}$.
In addition, we find the inertia correction factor $\FMA{\form_0}$ that vanishes only in the simple slab geometry.
On the right hand side of Eq.~\eqref{eq:perp_kinetic} we further find the Lorentz force term, which includes the geodesic transfer term. This term represents an energy transfer to the internal energy density Eq.~\eqref{eq:energy_pressure} since we know the Lorentz force to transfer angular momentum to the parallel angular momentum density. Disregarding the inertia correction factor $\form_0$ this term was also identified earlier to transfer energy to the zonal flow~\cite{Hallatschek2000,Scott2005,Naulin2005}. 

\begin{figure}[htbp]
    \centering
    \includegraphics[width = 0.5 \textwidth]{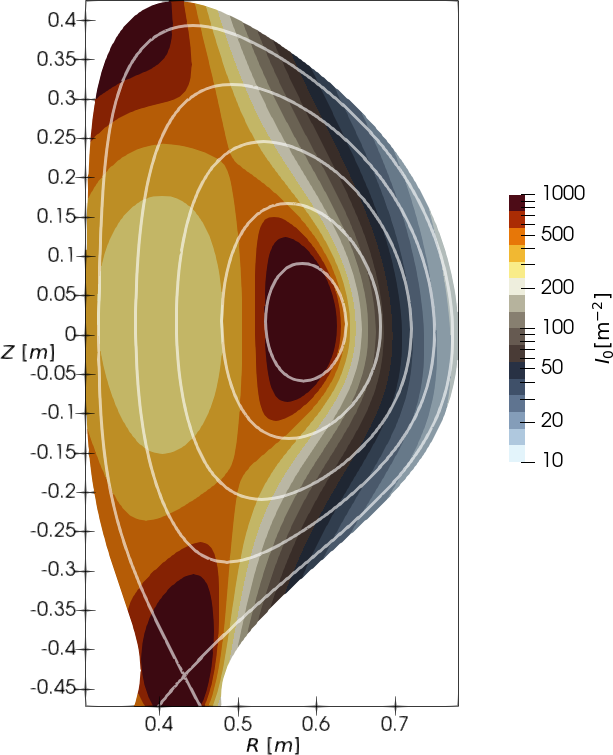}
    \caption{The $\form_0$ factor on an example tokamak equilibrium. Note the logarithmic colour scale, which is cut at the top at $1000$ due to the divergence at the X-point and the O-point. The contour lines are given at $\rho_t = \{0.2,0.4,0.6,0.8,1.0\}$ with the toroidal flux label $\rho_t := \sqrt{\psi_t/\psi_{t\mathrm{,sep}}}$.}
    \label{fig:h0square}
\end{figure}

\begin{figure}[htbp]
    \centering
    \includegraphics[width = 0.5 \textwidth]{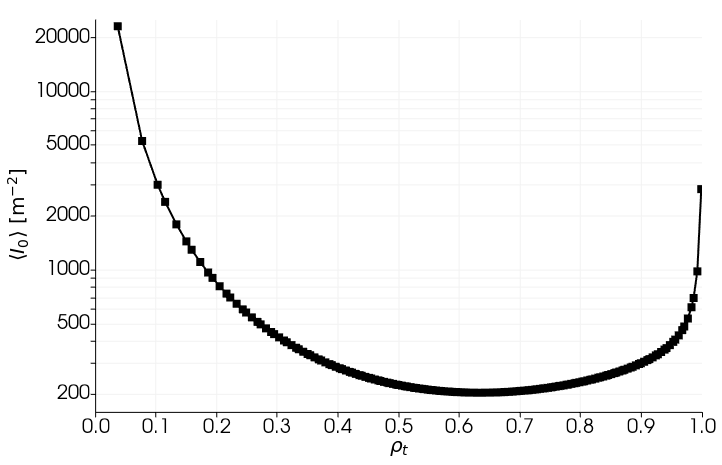}
    \caption{The $\RA{\form_0}$ factor on an example tokamak equilibrium as a function of the toroidal flux label $\rho_t := \sqrt{\psi_t/\psi_{t\mathrm{,sep}}}$. Note the logarithmic scale of the y-axis. }
    \label{fig:h0square_fsa}
\end{figure}

The second term on the right hand side is a novel term that appears for fluctuating radial velocity $\FF{u^v}$ and the inertia factor $\FF{\form_0}$.
In order to estimate the importance of the inertia factor we plot $\form_0$ for an exemplary tokamak equilibrium in Fig.~\ref{fig:h0square}.
We immediately see that the inertia factor
is not a flux function and is much smaller on the low-field side than on the high-field side. Furthermore, it diverges
at the X-point and the O-point. At the same time the toroidal component of the \ExB velocity $u_{E,\varphi}$ is zero at these points since the magnetic field is purely toroidal (and thus the zonal flow energy remains finite).
Further, the divergence at the X-point is an integrable singularity 
as shown in Fig.~\ref{fig:h0square_fsa}, where we plot the flux-surface average $\RA{\form_0}$.
Here, we mainly see that there appear
gradients close to the separatrix and in the core of the domain.

Finally, on the right hand side of Eq.~\eqref{eq:zonal_energy}
we find the source term $\mathcal S_{\mathrm{zonal}}$. This term
contains a contribution from the density source $S_n$ proportional to the inertia factor and the square toroidal \ExB velocity. The sign of this contribution depends only on the sign of $S_n$. Comparing to Fig.~\ref{fig:h0square} we see that the inertia factor is almost 2 orders of magnitude higher on the high field side than on the low field side. A particle source on the tokamak high field side is a far more effective source for zonal flow energy than on the low field side. This
supports experimental evidence that H-mode access is favored
by fueling plasma on the inboard side of a tokamak (for example in MAST~\cite{Akers2002}). A second contributor is the angular momentum source defined in Eq.~\eqref{eq:source_uEeta}, which we already discussed to be pronounced for poloidally asymmetric particle sources.

\section{Discussion}\label{sec:discussion}

\subsection{Simplified magnetic field geometries}\label{sec:simplified-geometries}
It is common in the existing literature to reduce the full three-dimensional magnetic field geometry to simplify expressions. The general magnetic field in Eq.~\eqref{eq:magvector} with both toroidal and poloidal components reduces to a purely toroidal magnetic field  for $\psi_p=0$ and the
purely poloidal field  for $\psi_t=0$. All our results so far hold for the general magnetic field without axisymmetry. We thus first discuss the poloidal and toroidal fields without assuming axisymmetry. A glance at the gyro-kinetic 1-form Eq.~\eqref{eq:spatial_oneform} convinces
us that in each of these cases both the poloidal and toroidal angular momentum have a single component. In a poloidal field the poloidal angular momentum contains only the parallel velocity $u_\parallel b_\vartheta$ while the toroidal angular momentum consists only of the \ExB flow $u_{E,\varphi}$ and vice versa for the purely toroidal magnetic field geometry. 

For the poloidal field the resulting evolution equations are actually already available and we do not need to compute anything further. The relevant equations are Eq.~\eqref{eq:reynolds_flow_fa} and the $\vartheta$ component of \eqref{eq:parallel_momentum_direction}.
For the purely toroidal magnetic field the parallel momentum balance is given by the $\varphi$ component of \eqref{eq:parallel_momentum_direction}, however the \ExB 
momentum is problematic since $\iota$ is zero and thus Eq.~\eqref{eq:reynolds_flow_vartheta_fa} does not hold.
Furthermore, since $\psi_p$ vanishes the flux-surface average needs to be redefined with the help of $\psi_t$.

In the following we will discuss the axisymmetric case for the general, the purely toroidal and the purely poloidal magnetic fields, which allows further simplifications.
\subsubsection{General axisymmetric magnetic field}\label{sec:axisymmetric}
An axisymmetric magnetic field can be written as in Eq.~\eqref{eq:symmetric_magnetic_field} and is a general feature of the tokamak configuration. It is well known that in this case the toroidal angular momentum density is a  conserved quantity~\cite{Scott2010,Brizard2011}. 
In our derivation axisymmetry leads to
the full toroidal angular momentum conservation (up to external sources)
in the $\varphi$ component of Eq.~\eqref{eq:angular_momentum_full}. The $\varphi$ derivatives in the first two terms on the right hand side vanish and the magnetic shear does not contribute. Comparing this result obtained in the \DK ordering
 to the exact result obtained using Noether's theorem~\cite{Scott2010}
we find a difference of half of the diamagnetic drift. The factor one half is difficult to interpret physically. In our derivation
we used the pressure equation  to evaluate this term and obtain the full diamagnetic drift. 
At the same time there is a freedom in how this term is treated in that we
could equally cast the diamagnetic drift completely under the time derivative instead of the right hand side. We comment more on this feature in Sec.~\ref{sec:emfields}.
\subsubsection{Purely toroidal, axisymmetric magnetic field}\label{sec:slab}
In the axisymmetric case we discuss here we can write (with cylindrical coordinates $R$, $Z$ and toroidal angle $\varphi$).
\begin{align}
   \psi_t &= \int^R B_0(R')\d R' ,\quad \psi_p = 0 \\
   \vec B(R) &= B_0\frac{R_0}{R} \ehat_\varphi \\
   \RA{h}_Z &:= \frac{1}{2\pi L_Z} \iint  h(R,Z,\varphi)\d Z\d \varphi
\end{align}
The gyro-kinetic 1-form Eq.~\eqref{eq:spatial_oneform} becomes $\gamma = (q\psi_t \d Z + m w_\parallel b_\varphi \d \varphi +m \mu \d \theta /q  $ and now
has symmetry in both the $R$ and $Z$-direction, 
which makes both $q \psi_t(R)$
and $mw_\parallel R$ conserved quantities separately. This is in fact an important point to emphasize. The purely toroidal magnetic field has two symmetries and thus two exactly conserved quantities instead of just one in the general axisymmetric geometry.
In the derivation of the poloidal \ExB momentum in Section~\ref{sec:main}, all we have to do 
is replace $\vn\psi_p$ with $\vn \psi_t(R) = B(R) \ehat_R$,
which defines $\vec \eta := \ehat_\varphi \times \vn \psi_t/B(R) \equiv \ehat_Z$.
Equation~\eqref{eq:reynolds_flow_vartheta_fa} thus reads (with zero magnetic shear)
\begin{align}
     \sumsp &\left\{m\RA{n}_Z\left(\frac{\partial}{\partial t} +  \mathcal U^R\frac{\partial}{\partial R}\right) \FA{u_{E,Z}}_Z
\right\}+\frac{\partial}{\partial R}\Mstress{Z}{\; R}
        \nonumber\\%
    =&  \sumsp\left\{-m\RA{n}_Z\left(\frac{\partial}{\partial R}  \Tstress{\perp,Z}{\;R} +  \Tstress{\perp,Z}{\;R}  \frac{\partial}{\partial R}\ln\RA{n}\right)+ m\mathcal S_{u_{E,Z}}\right\} 
 \label{eq:reynolds_flow_slab}
\end{align}
In the limit $B_0(R) = B_0$ and without $A_{1,\parallel}$ and finite Larmor radius effects this equation
agrees with~\cite{Held2018}.
The parallel angular momentum balance Eq.~\eqref{eq:parallel_momentum_direction_fa} now reduces to
\begin{align}\label{eq:parallel_momentum_slab}
    \sumsp&\left\{m\RA{n}_Z\left(\frac{\partial}{\partial t}
    +\mathcal U^R\frac{\partial}{\partial R}\right)\FA{u_{\parallel} R}_Z + \frac{\partial}{\partial R}\Kstress{\parallel,Z}{\;R}\right\}
    \nonumber\\
    = \sumsp&\left\{-m\RA{n}_Z\left(\frac{\partial}{\partial R}   \Tstress{\parallel,Z}{\;R} +  \Tstress{\parallel,Z}{\;R} \frac{\partial}{\partial R}  \ln\RA{n}_Z \right) + m\mathcal S_{u_\parallel R}\right.
\end{align}
Due to the symmetry in $R$ and $Z$ neither the Lorentz force, nor the mirror force appears in Eqs.~\eqref{eq:reynolds_flow_slab} and \eqref{eq:parallel_momentum_slab}. Further note that the continuity equation $\partial_t \RA{n}_Z + \partial_R (\mathcal U^R \RA{n}_Z ) = \RA{S_n}_Z$ can be used to cast these equations into conservative form.
\subsubsection{Purely poloidal, axisymmetric magnetic field}\label{sec:poloidal}
 The poloidal field approximation with $\psi_t = 0$ is potentially interesting for the field-reversed configuration~\cite{Binderbauer2015}, provided that our orderings in Section~\ref{sec:fundamentals} and \ref{sec:main} hold.
 We will here investigate the axisymmetric case since, as discussed before, the non-axisymmetric case is 
 already covered.
The Poincar\'e 1-form Eq.~\eqref{eq:spatial_oneform} reads
$
    \gamma = q \psi_p \d \varphi + mw_\parallel\sqrt{g_{\vartheta\vartheta}}^{-1}  \d \vartheta + \frac{m}{q}\mu \d \theta
$.
This results in $B_\parallel^* = \vec B^*\cdot \ehat_{\vartheta}=\vec B \cdot \ehat_{\vartheta}\equiv B_p$ with $\ehat_{\vartheta} := \vec e_\vartheta /|\vec e_\vartheta|$.
The approximation clearly breaks at the X-point where $B_{p}=0$,
however this point might
be redundant since flux coordinates themselves do not exist on the last closed
flux-surface where $\iota^{-1}$ diverges as we discussed in Section~\ref{sec:magnetic}.

It is interesting to note that toroidal symmetry now leads to the exact
conservation of $\gamma_{\varphi}=qA_{\varphi}=q\psi_p$ since $\ehat_{\vartheta}$ has no component in $\d\varphi$ in a symmetric situation. The toroidal angular momentum conservation in the poloidal field approximation thus contains only the toroidal component of the \ExB motion.
In this case we can write (note that Eq.~\eqref{eq:directional_varphi} still holds)
$
    \vec \eta := {\ehat_\vartheta\times\vn\psi_p}/{B_p}  = \ephi
$
which is possible with Eq.~\eqref{eq:directional_varphi},
$ \vec B = B_p \ehat_\vartheta $ and $\etheta\cdot \ephi = g_{\vartheta\varphi} = 0$. The vector $\vec\eta$ thus points in the
actual toroidal direction and does not have a poloidal component. We further have
$
        u_E^\tv = -\frac{\d v}{\d\psi_p} \frac{\partial\phi}{\partial\varphi}/B_p
        $
In comparison, we have that $u_{E,\vartheta} =0$, that is in the poloidal field approximation $\vec u_E = \ehat_\vartheta \times \nabla\phi/B_p$
has no poloidal component.
The non-zero part of the momentum fluxes is thus
\begin{align}\label{eq:reynolds_flow_poloidal}
    \sumsp &\left\{m\RA{n}\left(\frac{\partial}{\partial t} + \mathcal U^v\frac{\partial}{\partial v}\right) \FA{u_{E,\varphi}} \right\} +\frac{\partial}{\partial v} \Mstress{\varphi}{\tv}
        \nonumber\\%
    =&  \sumsp\left\{-m\RA{n}\left(\frac{\partial}{\partial v}  \Tstress{\perp,\varphi}{\tv} + \Tstress{\perp,\varphi}{\tv} \frac{\partial}{\partial v}\ln\RA{n}\right) + m\mathcal S_{u_{E,\varphi}}\right\} 
\end{align}
where we used that the $\varphi$ component of $\vec j_f \times \vec B$ vanishes with $((\ehat_\vartheta\times \vn\ln B_p) \times\ehat_\vartheta)_\varphi = ((\ehat_\vartheta\times \vec \kappa) \times\ehat_\vartheta)_\varphi  = 0$ due to the symmetry.
This means that in the poloidal field approximation
there is no transfer term between \ExB motion and parallel momentum just as in the purely toroidal magnetic field in Eq.~\eqref{eq:reynolds_flow_slab}.

In contrast the equation for the parallel momentum in toroidally symmetric cases becomes (with $b_\varphi =0$ and $b_\vartheta = \sqrt{g_{\vartheta\vartheta}}$)
\begin{align}\label{eq:parallel_momentum_poloidal}
    \sumsp&\left\{m\RA{n}\left(\frac{\partial}{\partial t}
    +\mathcal U^v\frac{\partial}{\partial v}\right)\FA{u_{\parallel}b_\vartheta} + \frac{\partial}{\partial v}\Kstress{\parallel,\vartheta}{\tv}\right\}
    \nonumber\\
    = \sumsp&\left\{-m\RA{n}\left(\frac{\partial}{\partial v}   \Tstress{\parallel,\vartheta}{\tv} +  \Tstress{\parallel,\vartheta}{\tv} \frac{\partial}{\partial v}  \ln\RA{n} \right) + m\mathcal S_{u_\parallel b_\vartheta}\right.
    \nonumber\\
   &-\left. \RA{ p_\perp \frac{\partial \ln B}{\partial \vartheta} +(p_\parallel + mnu_\parallel^2 )b^\vartheta \frac{\partial b_\vartheta}{\partial \vartheta}} \right\}
\end{align}
In contrast to the purely toroidal magnetic field here we find the mirror force and the geometric correction in the poloidal direction on the right hand side. This 
means that the background magnetic field acts as a source/sink of parallel momentum. 

Again, we note that the continuity equation $\partial_t \RA{n} + \partial_v (\mathcal U^v \RA{n} ) = \RA{S_n}$ can be used to cast Eqs.~\eqref{eq:reynolds_flow_poloidal} and \eqref{eq:parallel_momentum_poloidal} into conservative form.
\subsection{The momentum of electromagnetic fields in matter}\label{sec:emfields}
We now note that we have the possibility to rewrite
Eq.~\eqref{eq:reynolds_flow} using identity Eq.~\eqref{eq:onehalf} to cast the diamagnetic drift under the time derivative (using $u_{E}^\tv = -\np\phi\cdot \vec e_\varphi \d v/\d\psi_p$)
\begin{align} \label{eq:reynolds_flow_em}
    \frac{\partial }{\partial t}&\RA{\left(\vPem\times\vec B\right)_\varphi} \nonumber\\
    &- \frac{\partial}{\partial v}\RA{  E_\varphi \Pem^\tv + \left(\frac{1}{\mu_0}  B_{1,\perp,\varphi} -M^{\mathrm{em}}_{\perp,\varphi}\right) B_{1,\perp}^\tv }
    \nonumber\\
    &= -\RA{ (\vec j_f \times\vec B)_\varphi} + S_{\mathrm{em},\varphi}
\end{align}
with
\begin{align}
    \vec E &:= -\np \phi\\
    \vPem &:= -\sumsp m n \left( \frac{\np \phi}{B^2} + \frac{\np p_\perp }{qnB^2}\right)  \label{eq:pem}  \\
    \vMem &:= \sumsp m\frac{\bhat\times\vn (q_\parallel + u_\parallel p_\perp)}{qB^2} \label{eq:mem} \\
    \vec S_{\mathrm{em}} &:= \sumsp m S_n  \frac{\bhat\times\vn \phi}{B} + \frac{m\bhat\times\vn S_{p_\perp} }{qB} \label{eq:em_source}
\end{align}

Equation~\eqref{eq:reynolds_flow_em} is the evolution equation for the electromagnetic momentum flux $\vec g := \vec D \times \vec B$. The
electric part in the displacement field $\vec D := \epsilon_0\vec E + \vPem$ vanishes because we neglected the corresponding field part of
the action~\eqref{eq:action} and have quasineutrality. 
The momentum tensor has the form $T_{\varphi}^\tv:= -E_\varphi D^v - H_\varphi  B_{1,\perp}^\tv$ with the magnetizing field $\vec H  := {\vec  B_{1,\perp}}/\mu_0 - \vMem$.
The momentum flux $\vec g$ and tensor $T$ correspond to the ones given in Reference~\cite{Medina2017}.
With the identification
of the Lorentz force density $\vec f_L = \vec j_f \times \vec B$ on the right hand side we can write Eq.~\eqref{eq:reynolds_flow_em} as 
\begin{align}\label{eq:em_momentum}
\frac{\partial}{\partial t}\RA{g_\varphi} + \frac{\partial}{\partial v} \RA{T_{\varphi}^{\tv}} = - \RA{f_{L,\varphi}} + \RA{S_{{\mathrm{em},\varphi}}}
\end{align}
Notice the minus in the Lorentz force, which is a signature
that $g_\varphi$ is indeed the momentum
flux for the electromagnetic field rather than for the plasma itself. Furthermore, the form of the Lorentz force motivates
the identification of $\vec j_f$ as the \textit{free} current as opposed to the bound polarization current.

The $\vartheta$ component of the momentum flux follows by multiplying Eq.~\eqref{eq:em_momentum} with $\iota^{-1}$
\begin{align}\label{eq:em_momentum_theta}
\frac{\partial}{\partial t} \RA{g_\vartheta} + \frac{\partial}{\partial v} \RA{T_{\vartheta}^{\tv}}  = - \RA{f_{L,\vartheta}} - \RA{T_{\vartheta}^\tv}\frac{\partial}{\partial v}\ln \iota + \RA{S_{{\mathrm{em},\vartheta}}}
\end{align}
Here, notably a contribution from the magnetic shear appears on the right hand side as a coupling term to the external magnetic field.


In Eq.~\eqref{eq:pem} we define the electromagnetic polarization charge $\vPem$
analogous to the magnetization $\vMem$~\eqref{eq:mem} (which we repeat here for convenience) and different from the gyro-centre polarization charge $\vPgy$
by half the diamagnetic drift. We remark that neither of these quantities is uniquely defined. The form $\vPem$ and $\vMem$ highlights the physical origin of polarization and magnetization in gyro-kinetic models. Here, we can view the plasma as a collection of charged discs that can be magnetized
and polarized. The disc polarization $\vec \pi := m\bhat \times \vec{\dot X}/B$ stems from the drift velocities and reflects that due to the drifts the gyro-orbits are no longer closed~\cite{BrizardHahm,Brizard2013}. Macroscopically, in our model we have $\vPem = mn {\bhat \times (\vec u_E + \vec u_D)}/{B}$.
On the other side, the magnetization $\Mem$ contains the moving electric dipole contribution. An electric dipole $\vec \pi$ 
that moves with velocity $v_\parallel \bhat$ along the
magnetic field lines induces a magnetic moment $\vec\mu = \vec \pi \times  v_\parallel \bhat$.
However, we only find the diamagnetic part to the moving dipole contribution.
We are missing the contribution $mn u_\parallel\bhat \times \vn \phi / B^2$ since we neglected the corresponding
nonlinear coupling terms in the Hamiltonian~\eqref{eq:hamiltonian}.
\subsection{Comparison to drift-fluid models} \label{sec:driftfluid}
We note that Eq.~\eqref{eq:reynolds_flow_em} can also be viewed as a relation for 
the radial force density $\RA{(\vPem\times\vec B)_\varphi} = \sumsp \RA{\frac{m}{q B^2}\left( qn\np \phi + \np p\right)\cn\psi_p}$ where the force density $-qn\vec E_\perp + \np p$ appears inside the bracket on the right hand side. If the right hand side of Eq.~\eqref{eq:reynolds_flow_em} is zero, the radial pressure gradient and the radial electric field strength balance each other.
Alternatively, we can rewrite Eq.~\eqref{eq:reynolds_flow_em} as 
\begin{align}\label{eq:diamagnetic_evolution}
   \frac{\partial}{ \partial t}&\sumsp m\RA{n(u_{E,\varphi}+u_{D,\varphi})}
   \nonumber\\
   &+
    \frac{\partial}{\partial v} \left[\sumsp m\RA{n(u_{E,\varphi} + u_{D,\varphi} )} \FA{u_E^v}\right]    
    \nonumber\\
    &+
    \frac{\partial}{\partial v} \left[\sumsp m\RA{n}\left(\FExBstress{\varphi}{\tv} + \Fdiastress{\varphi}{\mathrm{T}\tv} \right) +\Mstress{\varphi}{\mathrm{T}\tv} \right]
    \nonumber\\
    =&-\RA{(\vec j_f\times\vec B)_\varphi} 
    \nonumber\\
    &+ \sumsp m\RA{ S_n u_{E,\varphi} +  \frac{\bhat\times\vn S_{p_\perp}}{qB}\cdot\ephi}
\end{align}
where the sum of \ExB and diamagnetic drift appear under the time derivative.
We point out that Eq.~\eqref{eq:diamagnetic_evolution} compares to Eq.~\eqref{eq:reynolds_flow_fa}
and is distinguished by the appearance of the transpose of the diamagnetic and Maxwell stresses and
the additional appearance of a pressure source on the right hand side in a form analogous to a diamagnetic drift term. This latter term appears through the use of the pressure equation in  bringing the diamagnetic drift under the time derivative. The $\vartheta$-component of Eq.~\eqref{eq:diamagnetic_evolution} is obtained by multiplying with $\iota^{-1}$.
 Eq.~\eqref{eq:diamagnetic_evolution} is also the form closest to the drift-fluid (generalized) vorticity equation~\cite{Smolyakov2000,Simakov2003}. To compare one needs to take the flux-surface average over the generalized vorticity equation and then integrate over the volume. This immediately allows the interpretation of Eq.~\eqref{eq:diamagnetic_evolution} as the volume integrated
equation for a divergence free current or a closed current loop. The Favre decomposition needs to be introduced in order to recover our stress terms.
Further, the momentum balance that results from integration of the ensemble averaged kinetic Vlasov equation also has a similar form to Eq.~\eqref{eq:diamagnetic_evolution} as seen for example in Reference~\cite{Sugama2011}. The difference is that we only recover the lower order \ExB and diamagnetic velocities instead of the full plasma velocity.

We point out that the pressure source (heating) on the right hand side of Eq.~\eqref{eq:diamagnetic_evolution} is not present in the drift-fluid generalized vorticity equation with plasma-neutral interactions~\cite{Thrysoe2018}. Further, our source
terms disagree with Reference~\cite{Simakov2003}, where a momentum source instead of a density or pressure source is presented.
The cause for these differences should be clarified in future work. In the present formulation the momentum source term in Eq.~\eqref{eq:diamagnetic_evolution} reflects (i) the presence of a formal kinetic source $S$ on the right hand side of the gyro-kinetic Valsov equation~\eqref{eq:vlasov} that is (ii) quasi-neutral under species summation Eq.~\eqref{eq:quasineutral_sources} and is (iii) transformed according to the gyro-centre transformation rules Eq.~\eqref{eq:particle_source}.  On the other hand the Stringer-Winsor spin-up term agrees with our results.

Finally, we emphasize that the evolution equation for the \ExB flow  Eq.~\eqref{eq:reynolds_flow}, the evolution for the electromagnetic field momentum Eq.~\eqref{eq:reynolds_flow_em} and the interpretation as a radial force density or the sum of \ExB and diamagnetic drifts in Eq.~\eqref{eq:diamagnetic_evolution} are completely equivalent views of the same result. In particular, physical arguments made with one of the three equations immediately translate into the other two.

\subsection{Relation to the ion orbit loss mechanism} \label{sec:ion-orbit-loss}
The ion orbit loss mechanism~\cite{Chang2002,Ku2018,Connor2000,deGrassie2009,Stoltzfus2012,Stoltzfus2019} refers to the idea that ion orbits close to the X-point end on the divertor target or the vessel wall and are thus lost to the confined plasma region. It is thought that the poloidal magnetic field close to the X-point is small such that the grad-B curvature drift velocity dominates over the parallel velocity making ions drift across the separatrix. 
This then generates a net flux of positive charge out of the confined region. In particle phase space the ions that are on a loss orbit are situated on a "loss-cone" encompassing ions with small parallel velocity and large perpendicular velocity / magnetic moment.
It is reported that the perpendicular kinetic energy of the loss cone reaches down to thermal energies~\cite{Chang2002}.

The ion orbit loss is often invoked in models explaining the L-H transition~\cite{Connor2000,Chang2002,Ku2018}, where it is thought that the outward current leaves a small region inside the separatrix negatively charged, which generates a strong radial electric field. This
field in turn drives a strong poloidal shear \ExB flow that then forms the transport barrier typical for the high confinement mode. 
On the other side the same idea is used to explain intrinsic toroidal rotation~\cite{deGrassie2009,Stoltzfus2012,Stoltzfus2019}, the observation that the plasma rotates toroidally without controlled external sources like NBI.  The main ingredient here is to assume that the rate by which ions enter loss orbits depends on the direction of their parallel velocity. This then generates an asymmetry between losses of so-called co- and counter-current ions. Since ions carry toroidal momentum, the preferential loss in one direction accelerates the plasma in the other.

Since our derivation of poloidal and toroidal angular momentum balance makes no assumption on the form of the distribution function $F$ (in particular it does not assume that $F$ is Maxwellian) and the particle orbits are retained via Eqs.~\eqref{eq:xdot} the ion orbit loss mechanism must consequentially be contained in our results. Here, we want to identify the relevant terms for both poloidal and toroidal rotation.

The net surface integrated current $\int_{\psi_p} \vec j\cdot \vec \dA = \RA{\vec j\cn v}$\footnote{Recall the definition of the flux-surface average Eq.~\eqref{eq:fsa_vol} to see that this is indeed the area integral} flowing through a flux-surface $\psi_p$, in particular the separatrix, by magnetic drifts is given by 
\begin{align} \label{eq:ion-orbit-loss-current}
\sumsp& \RA{  qn(\vec u_{\vn  B} + \vec u_\kappa )\cn v} = \RA{\vec j_f\cn v}
\nonumber\\
 =& \sumsp \RA{\left(\|\mu B\|\vec K_{\vn B}+\| mv_\parallel^2\|  \vec K_\kappa \right)\cn v}
 \nonumber\\
 =& \frac{\d v}{\d \psi_p}\sumsp\RA{ (p_\perp \vec K_{\vn B} + (p_\parallel+mnu_\parallel^2)\vec K_\kappa)\cn \psi_p}
\end{align}
where we inserted the definition of curvature and grad-B drifts Eqs.~\eqref{eq:curvature-drift} and \eqref{eq:grad-B-drift} and the velocity space moments to emphasize the origin of $\vec j_f$ as particle drifts. At this point recall again that $\RA{\vec j_f\cn v}\equiv \RA{(\vec j_f\times \vec B)_\varphi} \d v/\d \psi_p \equiv \RA{(\vec j_f\times \vec B)_\vartheta} \d v/\d \psi_t $ by virtue of Eqs.~\eqref{eq:directional_varphi} and \eqref{eq:directional_vartheta}. The term described in Eq.~\eqref{eq:ion-orbit-loss-current} is nothing but the Lorentz force term that appears in our momentum equations in Section~\ref{sec:momentum_fa} and which we already identified as the Stringer-Winsor spin-up or geodesic transfer term. 
The ion orbit loss contribution must be contained in the first term on the right side of Eq.~\eqref{eq:ion-orbit-loss-current} since it was argued that ions with large $\mu$ and small $v_\parallel$ fall on loss orbits. 
A signature of ion orbit loss would be if the ion term in Eq.~\eqref{eq:ion-orbit-loss-current} is larger than the electron contribution at or close to the separatrix.

At this point we notice that for favourable curvature drift direction
the curvature vectors counter-align with $\vn\psi_p$ ($\mathcal K(\psi_p) < 0$, decelerate) on the top 
and align ($\mathcal K(\psi_p) > 0$, accelerate) on the bottom of the tokamak. 
In order for the flux-surface average in Eq.~\eqref{eq:ion-orbit-loss-current} to yield a non-vanishing result we therefore need an up-down asymmetry of the pressure in the flux-surface. 
Furthermore, we notice that for our example tokamak equilibrium in Fig.~\ref{fig:flux_grid} we have $\RA{\vec K_\kappa \cn\psi_p}=0$. Indeed,
 more generally we find $\nc \vec K_{\vn B} = -\nc\vec K_\kappa = -\vec K_\kappa \cn\ln B ~\sim \mathcal O(\delta^6)$, 
 which results in $\partial_v \RA{K_{\kappa}} =  \partial_v (\RA{\vec K_\kappa \cn\psi_p}\d v  /\d \psi_p )\sim\mathcal O(\delta^6)$ and thus $\RA{\vec K_\kappa\cn\psi_p}\approx 0$.
 This means that only the fluctuations in $p_\perp$, $p_\parallel$ and $nu_\parallel^2$ contribute and we can 
 write\footnote{If we assume $p_\parallel = p_\perp = p$, we can further simplify
    $\RA{\vec j_f\cn\psi_p} 
    = \sumsp\left\langle\frac{\bhat\times\vn \RF{p}}{B}\cn \psi_p\right\rangle + \RA{m\RF{nu_\parallel^2}\vec K_\kappa\cdot\vn\psi_p}$
where we use that $\vec K_{\vn B} + \vec K_\kappa = \vec K$ and $\nc\vec K=0$ (see \ref{tab:operators}). Then we find the radial component of the diamagnetic drift $\RA{u_D^\tv}\d \psi_p/\d v$ in the first term on the  right hand side.}
 \begin{align}\label{eq:ion-orbit-loss-fluc}
     \RA{\vec j_f\cn\psi_p} = \sumsp\RA{ (\RF{p_\perp} \vec K_{\vn B} + (\RF{p_\parallel}+m\RF{nu_\parallel^2})\vec K_\kappa)\cn \psi_p}
 \end{align}

\subsubsection{Poloidal \texorpdfstring{\ExB}{TEXT} flow}
Even though, as argued in Eq.~\eqref{eq:angular_momentum_full} in Section~\ref{sec:momentum_fa}, the Lorentz force does not generate net poloidal momentum, it does generate \ExB momentum, respectively a radial electric field $u_{E,\vartheta}\sim \vn\phi\cn\psi_t$.
We thus conclude that the ion-orbit loss mechanism may indeed contribute to the radial electric field through the Lorentz force.

On the other hand, we emphasize that the Lorentz force is not the only candidate that contributes to the poloidal \ExB flow generation. Any other term in Eq.~\eqref{eq:reynolds_flow_vartheta_fa} could be equally important. Besides the \ExB Favre stress we identified for example the diamagnetic stress $\Fdiastress{\eta}{\tv}$ or the density gradient and magnetic shear related terms as additional candidates that may be equally relevant for the L-H transition.

\subsubsection{Intrinsic toroidal rotation}
The ion loss mechanism is through the Lorentz force indeed contained in the toroidal angular momentum conservation for \ExB~\eqref{eq:reynolds_flow_fa} and parallel \eqref{eq:parallel_momentum_direction_fa} angular momentum.
However, as we discussed in Eq.~\eqref{eq:angular_momentum_full} in Section~\ref{sec:momentum_fa} the Lorentz force does not actually generate net angular momentum, neither poloidal nor toroidal. A loss of ions through the separatrix does thus not generate toroidal angular momentum. As is shown in Eq.~\eqref{eq:angular_momentum_full} 
for an axisymmetric equilibrium the only sources for toroidal angular momentum are the actual source terms $\mathcal S_{nu_\parallel}$ and $\mathcal S_n$ on the right hand side. In order to explain an intrinsic rotation profile in this case we thus need to focus on the radial advection and stress terms, which describe the radial in-/outflow of momentum through the boundary flux-surface.
This requires a description of the turbulent fluctuations entering the stress terms, which is difficult to acquire
short of a full-scale simulation of the model equations.
The literature thus often invokes phenomenological models,
for example the asymmetric turbulent diffusion~\cite{Stoltzfus2012} where a preferential loss of co- or counter-current ions through the separatrix generates a net momentum gain for the remaining plasma inside the separatrix.

\subsection{Comparison to parallel acceleration}
The argument was made~\cite{Wang2013,Peng2017,Peng2017b,Wang2018} that in experimental measurements
the parallel velocity $u_\parallel$ is measured and not the parallel momentum density $nu_\parallel$. 
It was concluded that therefore $u_\parallel$ respectively $\RA{u_\parallel}$ should
be the quantity that theoretical work should
focus on when discussing intrinsic rotation.
In our view, neither premise nor conclusion of this hypothesis holds.
First, the velocity \textit{can} be measured at the same position and time as the density with for example velocity space tomography~\cite{Salewski2018} (and it should be noted that it is the velocity with respect to the line of sight rather than the parallel velocity that is actually measured in charge exchange diagnostics). Second, $u_\parallel$ is \textit{not} the angular momentum; $u_\parallel b_\varphi \approx u_\parallel R$ is and only part of it at that. Also, recall that even though it is not technically wrong to compute $\RA{u_\parallel}$ the flux-surface average~\eqref{eq:fsa_vol} is a volume average and should be taken over density like quantities (like $nu_\parallel$).
Finally, what comes out of a gyro-kinetic moment expansion (as performed in \cite{Wang2013,Peng2017,Peng2017b,Wang2018}) is the \textit{gyro-fluid} parallel velocity $U_\parallel$, not the actually measured fluid velocity $u_\parallel$.
As we discuss in Sec.~\ref{sec:gyro-fluid}
care must be taken when comparing gyro-fluid quantities like $U_\parallel$
to the actually physically measured fluid quantity $u_\parallel$ due to the involved coordinate transformation of Eq.~\eqref{eq:gyro-fluid-trafo}, which 
for $U_\parallel$ is given in Eq.~\eqref{eq:uparallel}.
The time evolution equation for $u_\parallel$ reads in our ordering (keeping terms up to $\mathcal O(\delta^3)$)
\begin{align} \label{eq:parallel_velocity}
    \frac{\partial u_\parallel}{\partial t} &
    + (\bhat + \bperp)\cn u_\parallel^2/2 + \vec u_E\cn u_\parallel 
    \nonumber\\
    &+ \frac{1}{mn}\nc\left((\bhat+\bperp) p_\parallel\right) + \frac{1}{m}t_\perp \npar\ln B 
    \nonumber\\
 &+ \frac{q}{m}\partial_t A_{1,\parallel} + \frac{q}{m}(\bhat+\bperp)\cn\phi = S_{u_\parallel}
\end{align}
The terms that appear beside the time derivative are in order the parallel advection term, the \ExB advection term, the parallel pressure gradient term, the mirror force term and the last two terms form the parallel electric field.
In Eq.~\eqref{eq:parallel_velocity} we see the local parallel
acceleration of a single (ion) species.
However,
working with accelerations instead of force densities as in Eq.~\eqref{eq:parallel_momentum_fa}
does not reveal that after species summation and flux-surface averaging
all net internal forces vanish and only external forces remain. As collectively generated, internal forces neither the pressure gradient nor the electric field can be the source of an intrinsic rotation profile. We point out here that the only external force that is able to make a contribution, the mirror force term $\RA{p_\perp \npar \ln B}$, was neglected in~\cite{Wang2013,Peng2017,Peng2017b,Wang2018}.



\section{Conclusions}
Our main results are the Favre averaged covariant poloidal and toroidal velocity evolution equations \eqref{eq:reynolds_flow_fa}, \eqref{eq:reynolds_flow_vartheta_fa}, and \eqref{eq:parallel_momentum_direction_fa} applicable in arbitrary magnetic field geometry including tokamaks, the reversed field pinch, the field-reversed configuration and stellarators. The \ExB equations~\eqref{eq:reynolds_flow_fa}, \eqref{eq:reynolds_flow_vartheta_fa} and the parallel components in Eq.~\eqref{eq:parallel_momentum_direction_fa} sum to the total angular momentum in Eq.~\eqref{eq:angular_momentum_full}.

The usefulness of the Favre-average formalism mainly stems from the identification of the Favre stress as the mediator between turbulent fluctuations and flux-surface averaged profiles.
In our full-F gyro-kinetic formulation the perpendicular Favre stress $\Tstress{\perp,\eta}{\tv}$ appears in the \ExB part of the angular momentum as a natural extension of the Reynolds stress through the density weighted flux-surface average - the Favre average~\cite{Held2018}. 
The perpendicular Favre stress consists of the previously found \ExB contribution $\FExBstress{\eta}{\tv}$, but also of the novel diamagnetic $\Fdiastress{\eta}{\tv}$ and magnetic flutter $\Xstress{\eta}{\tv}$ contributions defined in Eq.~\eqref{eq:exb_stresses}. Besides the Favre stress, the vacuum Maxwell stress $\MBstress{\eta}{\tv}$ and magnetization stresses $\Wstress{\eta}{\tv}$ defined in Eq.~\eqref{eq:maxwell_stress} appear. We highlight the relation to the general density gradient drive term in Eq.~\eqref{eq:derivativeTotalStress}.
 Furthermore, the Lorentz force originating from the curvature and grad-B drift induced currents represents a source for \ExB angular momentum density. 
Finally, poloidally asymmetric density sources Eq.~\eqref{eq:source_uEeta} contribute to angular momentum generation.

Analogous to the \ExB part, the parallel component of the angular momentum density Eq.~\eqref{eq:parallel_momentum_direction_fa} is generated by the parallel Favre stress $\Tstress{\parallel,\eta}{\tv}$ in Eq.~\eqref{eq:parallelfavrestress} as well as the kinetic stress $\Kstress{\parallel,\eta}{\tv}$ Eq.~\eqref{eq:kinetic_stress} stemming from magnetic fluctuations. The Lorentz force appears with an opposite sign as in the \ExB equation thus vanishing in the summed total angular momentum density in Eq.~\eqref{eq:angular_momentum_full}, both toroidally as well as poloidally. In addition, in Eq.~\eqref{eq:parallel_momentum_direction_fa} the mirror force appears as a source of parallel momentum density.  

We construct the inertia tensor from the first fundamental form in Eq.~\eqref{eq:fundamental}. The relevant discussion is based on the mean flow generated by the covariant, Favre averaged velocity components that we investigate in the first part of the paper. From there we construct the rotational energy in Eq.~\eqref{eq:rotational_energy}. The \ExB part of this energy can be split into a mean "zonal" and fluctuating part and we present the evolution of the mean in Eq.~\eqref{eq:perp_kinetic} using the previously derived evolution equations for angular momentum. The main finding compared to a simplified geometry is the appearance of a correction factor due to the inertia tensor, which in particular modifies the effect of the density source on the right hand side. A density source on the high field side is a more effective source of zonal flow energy than on the low field side.

We show that we recover previous results obtained in simplified geometries. Interestingly, the purely toroidal magnetic field leads to the exact conservation of both the poloidal \ExB velocity as well as the parallel angular momentum density. This is because an additional symmetry is introduced by this geometry. 
We also point out that the ion orbit loss mechanism as outlined in the literature is identical to
the "geodesic transfer term" and the "Stringer-Winsor spin-up mechanism" and is contained in our results in the Lorentz force term on the right hand side of the poloidal \ExB angular momentum equation~\eqref{eq:reynolds_flow_vartheta_fa}.
Finally, we clarify several misconceptions in connection with "parallel acceleration" relating previous findings to our results.

The main drawback of our derivation is the \LWL\ in the gyro-kinetic action Eq.~\eqref{eq:action}, which 
effectively reduces our model to a drift-kinetic model and misses
higher order finite Larmor radius and polarization effects that could play a role for the L-H transition. 
We mainly perform this limit in order to avoid an infinite sum in the relation between the ordinary and the variational derivative Eq.~\eqref{eq:derivative_as_variation} and to avoid the introduction of a fluid closure of the infinite expansions in the polarization and gyro-averages~\cite{Held2020}. The \DK ordering in Section~\ref{sec:main} avoids geometrical correction factors
stemming from for example perpendicular derivatives on the magnetic field unit vector in Section~\ref{sec:main} and allows to recover fluid (as opposed to gyro-fluid) moments in our equations and to compare to existing drift-fluid models via Eq.~\eqref{eq:diamagnetic_evolution}. However, our momentum balance equations are only valid up to order three and the energy balance equations up to order four within this ordering. Future work could address the above issues.

Our results can be used to verify simulation results.
The application of these results within full-F gyro-fluid models is subject of ongoing research.
However, as previously stated, the available equations in this work are by no means restricted to gyro-fluid models since the derivation contains no assumption  on the form of the distribution function. Thus the presented results are relevant also for other frameworks beyond gyro-fluid models like for example gyro-kinetic or drift-fluid models.

The experimental validation of our results may be challenging due to the number of terms that appear in the evolution equations \eqref{eq:reynolds_flow_fa}, \eqref{eq:reynolds_flow_vartheta_fa}, and \eqref{eq:parallel_momentum_direction_fa} that in particular require the measurement of plasma potential, parallel velocity, density, pressure and possibly magnetic field fluctuations at the same time and positions. Further, a problematic operation is the flux-surface average. The argument that a time average over the measurement interval equates
the flux-surface average only holds if the measured quantity is a flux-function in the first place. On the other hand we provide the theoretical foundation for a discussion of the dominant physical mechanisms that 
generate poloidal and toroidal angular momentum density and rotational energy in any toroidal magnetic field configuration.

\paragraph{Acknowledgements}
We acknowledge fruitful discussions with N. Tronko, P. Strand, V. Naulin, and J.J. Rasmussen. The research leading to these results has received funding from
the European Union’s Horizon 2020 research and innovation programme under the Marie Sklodowska-Curie grant agreement no.
713683 (COFUNDfellowsDTU). This work has been carried out within the framework of the EUROfusion Consortium and has received funding from the Euratom research and training programme 2014-2018 and 2019-2020 under grant agreement No 633053. The views and opinions expressed herein do not necessarily reflect those of the European Commission.

\appendix

\section{Formulary}\label{app:formulary}
\begin{table*}[htbp]
\caption{Definitions of geometric operators with $b^i$ the contra-variant components of $\bhat$ and $g^{ij}$ the contra-variant elements of the metric tensor. We assume $(\vn\times\bhat)_\parallel = 0$. 
}\label{tab:operators}
\centering
\begin{tabular}{lll}
\textbf{Name} &  \textbf{Symbol} & \textbf{Definition} \\
\midrule
    Projection Tensor&
    $h $ & $h^{ij} := g^{ij} - b^ib^j $   \quad \text{ Note }$ h^2=h$\\
    Perpendicular Gradient&
    $\np $&
    $ \np f := \bhat\times(\vn f\times \bhat ) \equiv
    h \cdot \vn f$ \\
        Perpendicular Divergence&
    $\np^\dagger $&
    $ \np^\dagger \cdot \vec v := ´-\nc( h \cdot \vec v) = -\nc\vec v_\perp$ \\
    Perpendicular Laplacian &
    $\Delta_\perp $ &
    $ \Delta_\perp f:= \nc (\np f)
    = \nc( h\cdot\vn f) \equiv -\np^\dagger\cdot\np$  \\
    Curl-b Curvature Operator &
    $\mathcal K_{\kappa}$ &
    $\mathcal K_{\kappa}(f) := \vec{ K_{\kappa} }\cn f = \frac{1}{B}(\bhat\times\vec\kappa)\cn f$ \quad \text{ with }$\vec\kappa :=\bhat\cdot\vn\bhat$ \\[4pt]
    Grad-B Curvature Operator &
    $\mathcal K_{\vn B} $ &
    $\mathcal K_{\vn B}(f) := \vec{ K_{\vn B}} \cn f = \frac{1}{B}(\bhat \times\vn \ln B)\cn f$ \\[4pt]
    Curvature Operator&
    $\mathcal K$ &
    $\mathcal{K}(f):=\vec{ K} \cn f =
     \vec{\vn}\cdot\left(\frac{\bhat\times\vec{\vn} f}{B}\right) =\vn \times \frac{\bhat}{B} \cn f$,\\[4pt]
    Parallel derivative&
    $\npar $&
    $ \npar f := \vec B\cdot\vn f/B$ \quad  Notice $\nc\bhat = -\npar\ln B$ \\
\bottomrule
\end{tabular}
\end{table*}
\subsection{Flux surface and Favre average}
The flux-surface average (see for example \cite{haeseleer}) is an average over a
small volume - a differential shell centered around the flux-surface.
We define
\begin{align} 
\langle f \rangle (\psi_p) :=& \frac{\partial}{\partial v} \int_\Omega \dV f 
 = \int_{\psi_p}  \frac{f(\vec x)}{|\vn v|} \dA
 \end{align}
 where we define $v(\psi_p) := \int_0^{\psi_p} \dV$ as the volume
flux label
and for the second identity, recall the co-area formula
\begin{align} \label{eq:coarea}
\int_{\Omega} f(\vec x) \dV =
\int_0^{\rho} \d \rho' \left( \int_{\rho'=\mathrm{const}}    \frac{f(\vec x)}{|\vn \rho|} \dA\right) 
\end{align}
where $\rho(\psi_p)$ is any flux label and $\Omega$ is the volume enclosed by the contour $\rho=\mathrm{const}$. In flux coordinates we have $\dA = \sqrt{g} |\vn\rho| \d\vartheta\d\varphi$.
The co-area formula can be viewed as a change of variables in the
volume integral.
The average fulfills the identities (with scalars $\lambda$ and $\mu$)
\begin{align}
\langle \mu f + \lambda g\rangle &= \mu\langle f\rangle + \lambda \langle g\rangle \\
\langle f(\psi_p) g(\vec x) \rangle &= f(\psi_p) \langle g(\vec x)\rangle \\
\RA{ \nc \vec j} &= \frac{\partial}{
  \partial v} \RA{\vec j \cdot \vn v} 
  \nonumber\\
  &= \left(\frac{\d v}{\d \rho}\right)^{-1}
  \frac{\partial}{
  \partial \rho} \left(\frac{\d v}{\d\rho} \RA{\vec j \cdot \vn \rho}\right)
\end{align}
Also note that for any divergence free vector field $\nc \vec j = 0$ and a flux function $f(\psi_p)$ we have
\begin{align}
    \langle \nc( \vec j f)\rangle = 0
\end{align}
which is proven straightforwardly.

We note the Reynolds decomposition for any function $h(\vec x)$
\begin{align}
    h &\equiv \RA{h} + \RF{ h}
\end{align}
and its generalization, the Favre average and decomposition
\begin{align}
    \FA{h} &:= \frac{\RA{nh}}{\RA{n}} \\
    h &\equiv \FA{h} + \FF{ h}
\end{align}
where $n$ is the particle density, which makes the Favre average species dependent.
The Favre average fulfills
\begin{align}\label{eq:FA_identities}
    \FA{ gh} &= \FA{g}\FA{h} + \FA{\hat g\hat h}
\end{align}
It is sometimes useful to define the Favre average using the total mass density $\rho_M :=\sumsp mn$ as
\begin{align}
    \FMA{h} := \RA{\rho_M}^{-1}\sumsp m \RA{nh}
\end{align}
for any (possibly species dependent) function $h$. If $h$ is species independent this definition simplifies to $\FMA{h}=\RA{\rho_M h}/\RA{\rho_M}$.

\subsection{Fluid moments}\label{sec:fluid_moments}
The velocity space moments Eqs.~\eqref{eq:fluid_moments} and \eqref{eq:source_moments} read
\begin{align*}
    \|\zeta\| &:=\int \d w_\parallel \d \mu m^2 B F\zeta  \\
    \|\zeta\|_S &:= \int \d w_\parallel \d \mu\d \theta m^2 B S\zeta
\end{align*}
 \begin{table}[hbtp]
\begin{center}
\begin{tabular}{llll}
\hline
$N$  & $\|1\|$ & 
$ P_\perp\equiv NT_\perp$ & $\|\mu B\| $ \\
 $NW_\parallel$ & $\| w_\parallel\|$
 &
$P_\parallel \equiv NT_\parallel$ & $\| m(w_\parallel - W_\parallel)^2\|$ \\
 $ N\phi$ & $ \|\phi\|$ &
$ Q_\parallel$  & $\|\mu B(w_\parallel -W_\parallel)\|$  \\ 
 $N A_{1,\parallel} $ & $\| A_{1,\parallel}\|$ & 
 \\
\hline
\end{tabular}
\end{center}
\caption{List of the first few gyro-fluid moments: gyro-fluid density $N$, parallel canonical velocity $W_\parallel$ the perpendicular and parallel
pressure ($P_\perp$ and $P_\parallel$)/ temperature $T_\perp$ and $T_\parallel$ as well the parallel heat flux $Q_\parallel$.}
\label{tab:gyro-fluid}
\end{table}
In Table~\eqref{tab:gyro-fluid} we name the first few velocity space moments of the gyro-kinetic distribution function $F$.
The moments over the gyro-kinetic source function $S$ are named
analogous as $S_N$, $S_{NW_\parallel}$, $S_{P_\perp}$, $S_{P_\parallel}$ and $S_{Q_\parallel}$.
We can identify $ \|mv_\parallel\| =  mNW_\parallel - q N\mathcal A_{f,\parallel} \equiv mNU_\parallel$ and
\begin{align}
\|mv_\parallel^2\| &= P_\parallel + mNU_\parallel^2,\quad
\| \mu B v_\parallel \|= Q_\parallel + P_\perp U_\parallel, 
\label{eq:gyro-fluid}
\end{align}
The relation between gyro-fluid quantities $N(\vec X, t)$, $U_\parallel(\vec X, t)$, ... given in gyro-centre coordinates $\vec X$
and the physical fluid quantities, which we denote with lower case letters $n(\vec x,t):= \int\d^3v f(\vec x, \vec v, t)$, $u_\parallel(\vec x,t) := \int \d^3 v v_\parallel f(\vec x,\vec v, t)$ ..., where $f(\vec x,\vec v,t)$ is the distribution function in
particle phase-space (and we here overburden the use of $v$ as the velocity instead of the volume flux-label) is
given by Eq.~\eqref{eq:gyro-fluid-trafo}
\begin{align}
    \|\xi\|_\vec{v} = \|\zeta\| + \Delta_\perp \left( \frac{m\|\mu B\zeta\|}{2qB^2}\right) + \nc\left( \frac{m \|\zeta\| \vn_\perp \phi}{B^2}\right)
\end{align}
This relation can be inverted
up to order $\delta^2$ as for example in Eqs.~\eqref{eq:particle_density} and \eqref{eq:uparallel}
\begin{align} 
    N &= n - \Delta_\perp \left(\frac{ mnt_\perp}{2q^2B^2}\right) - \nc \left( \frac{ mn }{qB^2} \np \phi\right)
    \\
    NU_\parallel &= nu_\parallel - \Delta_\perp \left( \frac{m( q_\parallel + u_\parallel p_\perp)}{2q^2B^2}\right)
\end{align}
Analogous relations hold for the moments of the gyro-kinetic source function $S_N$ and $S_{NU_\parallel}$.

\subsection{Fluid velocities}\label{sec:fluid_velocities}
We introduce for any vector $\vec u$
\begin{align}
u^v &:= \vec u \cn v  &&
\vn v := \frac{\d v}{\d \psi_p}\vn \psi_p \\
u_\varphi &:= \vec u \cdot \vec e_\varphi&&
u_\vartheta := \vec u\cdot \vec e_\vartheta &&
u_\parallel := \vec u\cdot \bhat
\end{align}
We define the \ExB drift $\vec u_E$, the
grad-B drift $\vec u_{\vn B}$, the diamagnetic drift $\vec u_D$, the curvature drift $\vec u_\kappa$, the first order magnetic fluctuations $\vec B_{1,\perp}$ and  the electromagnetic magnetization density $\Mem$
\begin{align}
\vec u_E &:= \frac{\bhat\times\vn \phi}{B}  
&& 
\vec u_{\vn B} := t_\perp \frac{\bhat\times\vn \ln B}{qB}\label{eq:grad-B-drift}
\\
\vec u_{D} &:=\frac{\bhat\times\vn p_\perp}{qnB}  
&&
\vec u_\kappa := (t_\parallel + mu_\parallel^2) \frac{\bhat\times\vec \kappa}{qB}\label{eq:curvature-drift}
 \\
 {\vec B}_{1,\perp} &:= {\vn A_{1,\parallel} \times\bhat}
 &&
\vMem :=\sumsp\frac{m\bhat\times \vn (q_\parallel + p_\perp u_\parallel)}{q B^2} \label{eq:emmagnetization}
\end{align}
Note that $\RA{\RF{ B_{1,\perp}^v}} =\mathcal O(\delta^3)$ in the \DK ordering and $\vec b_{1,\perp} := \vec B_{1,\perp}/B$.
Finally, we have the free current
\begin{align}\label{eq:free_current}
\vec j_f := \sumsp qn(\vec u_\kappa + \vec u_{\vn  B} )
\end{align}
originating in the particle curvature and grad-B drifts.


\bibliography{theory}
\bibliographystyle{iopart-num.bst}
\end{document}